\documentclass[epj]{svjour}
\usepackage{graphics}
\usepackage{graphicx}
\usepackage[square,sort,comma,numbers]{natbib}
\usepackage{xcolor}
\usepackage{amsmath}
\usepackage{txfonts}
\def\msun{\,\rm M_\odot}
\begin{document}
\title{Electromagnetic counterparts of black hole-neutron star mergers: dependence on the neutron star properties}
\subtitle{}
\author{C. Barbieri\inst{1,3}\thanks{\emph{e-mail:} c.barbieri@campus.unimib.it} \and O.~S.~Salafia\inst{2,3} \and A.~Perego\inst{3,4} \and M.~Colpi\inst{1,3} \and G.~Ghirlanda\inst{2,3}
%
}                     
\offprints{}          
\authorrunning{C.~Barbieri et al.}
\titlerunning{EM counterparts of BHNS mergers: dependence on the NS properties}

\institute{Universit\`a degli Studi di Milano-Bicocca, Dipartimento di Fisica ``G. Occhialini'', Piazza della Scienza 3, I-20126 Milano, Italy\label{unimib}\and INAF -- Osservatorio Astronomico di Brera, via E. Bianchi 46, I-23807 Merate, Italy\label{oab.me}  \and INFN -- Sezione di Milano-Bicocca, Piazza della Scienza 3, I-20126 Milano, Italy\label{infn.mib} \and Universit\`a degli Studi di Trento, Dipartimento di Fisica, via Sommarive 14, I-38123 Trento, Italy \label{unitn}}
\date{Accepted: 29 Oct 2019}
%
\abstract{
Detections of gravitational waves (GWs) may soon uncover the signal from the coalescence of a  black hole - neutron star (BHNS) binary, that is expected to be accompanied by an electromagnetic (EM) signal.
In this paper, we present a composite semi-analytical model to predict the properties of the expected EM counterpart from BHNS mergers, focusing on the kilonova emission and on the gamma-ray burst afterglow.
Four main parameters rule the properties of the EM emission: the NS mass $M_\mathrm{NS}$, its tidal deformability $\Lambda_\mathrm{NS}$, the BH mass and spin. Only for certain combinations of these parameters an EM counterpart is produced. 
Here we explore the parameter space, and construct light curves, analysing the dependence of the EM emission on the NS mass and tidal deformability. Exploring the NS parameter space limiting to $M_\mathrm{NS}-\Lambda_\mathrm{NS}$ pairs described by a physically motivated equations of state (EoS), we find that the brightest EM counterparts are produced in binaries with low mass NSs (fixing the BH properties and the EoS).
Using constraints on the NS EoS from GW170817, our modeling shows that the emission falls in a narrow range of absolute magnitudes. Within the range of explored parameters, light curves and peak times are not dissimilar to those from NSNS mergers, except in the B band. The lack of an hyper/supra-massive NS in BHNS coalescences causes a dimming of the blue kilonova emission in absence of the neutrino interaction with the ejecta.
\PACS{
      {97.60.Jd}{Neutron stars}\and
      {97.60.Lf}{Black holes}\and
      {04.30.−w}{Gravitational waves}\and
      {98.70.Rz}{Gamma-ray sources; Gamma-ray bursts}\and
      {26.30.−k}{Nucleosynthesis in novae, supernovae, and other explosive environments}
     } 
} 
\maketitle

\section{Introduction}

Starting from 2015 and during the first two observing runs (O1 and O2), Advanced LIGO and Advanced Virgo detected the gravitational wave (GW) signal from the coalescence of several black hole-black hole (BHBH) binaries \citep{Abbott18-10-bh} and of a neutron star-neutron star (NSNS) binary \citep{GW170817}.
A binary composed of a black hole and a neutron star (BHNS) is yet to be discovered \citep{Abadie2010}, but prospects for the first detection during the current LVC\footnote{LVC is the achronym of the LIGO Scientific Collaboration and Virgo Collaboration} observing run (O3) are encouraging.   

The BHNS merger rate prediction from population synthesis models is in the range [$\sim 10^{-9}$, $\sim 10^{-6}$] Mpc$^{-3}$  yr$^{-1}$ \citep{Abadie10,Clark2015,Dominik2015,Mapelli2018,Giacobbo2018}.
From the non-detection of such an event during O1 and O2, it is possible to draw a 90\% upper limit on the BHNS merger rate of $6.1 \times 10^{-7}$ Mpc$^{-3}$  yr$^{-1}$, assuming a black hole (BH)  mass of $M_\mathrm{BH}\sim5\msun$ with BH spins isotropically distributed, and a neutron star (NS) mass of $M_\mathrm{NS}\sim1.4\msun$ \citep{CatalogoGW}.

In O3 the sensitivity has been improved, extending the detection range. Therefore, a larger volume of the universe can be explored and, unless the true BHNS merger rate falls in the lower end of the estimates, we can expect to detect this kind of event for the first time ever.
For example, \cite{Dominik2015} estimate a BHNS merger detection rate for a 3-detector-network in O3 in the range [$\sim0.04$, $\sim12$] yr$^{-1}$ (depending on the assumed SNR threshold and on the metallicity evolution model).

The aim of this study is at exploring the dependence of the potential, prospected electromagnetic (EM) counterparts of BHNS mergers on the NS properties, extending the analysis presented in \cite{Barbieri2019}.

The NS mass is a key parameter in our analysis. 
Current observations of Galactic binary systems where at least one of the components is a NS \citep{Antoniadis2013,Ozel2012,Ozel2016,Antoniadis2016,Tauris2016,Tauris2017} indicate a mass range from $\sim 1.17\pm0.01 \msun$ to $\sim 2.01\pm 0.04 \msun$. \footnote{A much higher, albeit uncertain, mass of $2.7\pm0.21\msun$ is observed in the recycled millisecond pulsar J1748-2021B \citep{Freire2007}. However, due to uncertainties in the  assumed binary orbital inclination angle,
for J1748-2021B the authors indicate that there is $1\%$ probability that the NS mass is below $2~\msun$.}
The NSs in double-NS systems have masses between $1.174\msun$ and $1.559\msun,$ with uncertainties $\lesssim 0.01\msun$ \citep{Tauris2017}.
In the case of recycled NS with white dwarf companions, the mass distribution is broader and can be fitted by a bimodal
distribution with a low-mass component centered at $1.393^{+0.031}_{-0.029}\msun$ and a high-mass component with mean $1.807^{+0.081}_{-0.132}~\msun$ \citep{Antoniadis2016}. 
The most massive NS with an accurate and highly reliable mass estimate is the radio pulsar J0740+6620 in a low-mass binary, whose mass is $2.14^{+0.10}_{-0.09}\msun$ \citep{Cromartie2019}.
The theoretical maximum mass of non-rotating NSs is highly debated, being strongly dependent on the NS composition and equation of state (EoS). Interestingly, based on the joint GW and EM analysis of GW170817, \cite{Margalit2017} infer an upper limit on the maximum mass of a NS of $2.17\msun$ at 90\% confidence level.

While in this work we focus on the NS properties and their deformability, we still need to specify the BH mass and spin\footnote{Hereafter, the term spin refers to the dimensionless spin parameter $\chi_\mathrm{BH}=cJ/GM_\mathrm{BH}^2$, where $J$ is the angular momentum of the BH.} in order to make definite predictions. 
The masses of BHs observed in Galactic X-ray binaries cluster around a mean of $7.8\pm1.2\msun$  \citep{Ozel2010}. By contrast, those detected by the LVC have a broader distribution between $7.6^{+1.3}_{-2.1}\msun$ and $50.6^{+16.6}_{-10.2}\msun$ \citep{Abbott18-10-bh}.

Concerning BH rotation, as shown in e.g.~\cite{Shibata11} and discussed in \cite{Barbieri2019}, the BH spin (as also its orientation with respect to the binary angular momentum, the tilt angle) plays a key role in determining the properties of the EM counterparts of BHNS events. As explained in \cite{Belczynski2017} the effect of spins in GW waveforms is subdominant. Indeed the waveforms are most sensitive to the effective spin of the binary, which is a mass-weighted combination of the binary component spins \footnote{In this work we assume the NS spin to be negligible. Indeed the time delay between the NS formation and the binary merger is long enough so that the NS spin (initially large) decreases through dipole emission. Furthermore, the absence of matter accretion onto the NS avoids the spin-up through recycling. Thus the NS spin is expected to be low before tidal locking it remains negligible because the GW driven inspiral time is much shorter than the timescale for tidal spin-up \citep{Kochanek1992,Bildsten1992}.} along the binary angular momentum. The LVC detections of ten BHBH mergers indicate that the effective spins $\chi_{\rm eff}$ cluster around $\chi_{\rm eff} \sim 0$, while the individual BH spins are poorly constrained. \cite{Farr2018} proposed a model-agnostic approach to characterise the spin properties of the LVC BHBH population, showing that a strictly aligned formation scenario for the BH spins would imply that most individual spins take values $\lesssim 0.3$.
Recently \cite{Belczynski2017} noted that, based on the Geneva stellar evolutionary code \citep{Georgy2014} and on estimates on the CO cores prior the core collapse, BH natal spins can be as high as 0.85 for CO cores of less than $16\msun$, collapsing to form a BH of $\lesssim 20\msun$ \citep{ArcaSedda2019}. It is clear that forthcoming new observations will shed light on the natal spins of stellar BHs, challenging  theoretical models. 

As a final remark, we recall that the BH spins in Galactic binaries have been inferred using the continuum fitting of the thermal spectrum from the Novikov and Thorne relativistic thin-disc model \citep{Novikov1973}. The BHs observed in our Galaxy are either persistently X-ray bright sources (as Cyg X-1), or transient. The former carry spin values $>0.85$ with uncertainties at ten per cent level, and are natal in origin, while the latter have a large spread extending from $ \approx 0$ to $>0.95$ \citep{McClintock2014}. 
 Cyg X-1 will soon encounter a Roche lobe overflow episode, followed shortly by a Type Ib/c supernova and the formation of NS.
 According to evolutionary calculation by \cite{Belczynski2011,Belczynski2012-archive} only $\sim 1\%$ of Cyg X-1 like systems end forming a BHNS merging binary, the remaining ones either disrupting or never coalescing as mass transfer prior NS formation widens the orbit.
 Interestingly, Ultra Luminous X-ray sources could be the progenitors of BHNS systems, that formed through chemically homogeneous evolution \citep{Marchant2017}. These systems likely host massive and highly spinning BH, contrary to common envelope evolution systems for which the BH mass distribution appears to have in most cases a peak at around $M_{\rm BH}\sim 10\msun$ with an extended tail, as shown in \cite{Giacobbo2018}. It is clear that discovering BHNS merging systems both in GWs and through EM-emission will inform on the formation channels providing rather stringent tests on the models proposed. 

As an additional point, we recall that observations of Galactic compact binaries show a mass gap between the highest NS mass observed ($\sim 2\,M_\odot$) and the lowest BH mass ($\gtrsim 5\msun$) \citep{Ozel2010,Farr2011}. The compact object mass distribution is strictly related to the process of stellar core collapse and to the physical conditions in supernova (SN) explosions. It has been suggested that rapid SN explosions could produce this mass gap, while from delayed explosions a continuum mass spectrum for the remnants could arise \citep{Belczynski2012,Fryer2012}.

Binary population synthesis performed in \cite{Dominik2012} yields for merging BHNS a BH mass range [$\sim5$, $\sim40$] $\msun$, and a NS mass range [$\sim1.1$, $\sim2$] $\msun$, both depending on the models for stellar evolution and binary interaction. \cite{Giacobbo2018}, in their population synthesis performed using the \texttt{MOBSE} code \citep{Giacobbo2018_2}, found that for merging binaries the NSs are mostly massive ($1.3\lesssim M_\mathrm{NS}\lesssim2$ $\msun$) and the BHs have preferentially low masses ($5\lesssim M_\mathrm{BH}\lesssim 15 \msun$, with the lower value imposed ab initio depicting rapid SN explosion). 

As discussed in \S \ref{sec:bh_properties}, we select two values for the BH spin and two values for its mass, in order to bracket a range of possible configurations.

In a BHNS merger the final remnant is always a BH. However, General Relativity (GR) numerical simulations indicate that there are two possible fates for the NS, depending on the relative position of the BH innermost stable circular orbit (ISCO) and the distance at which the BH gravitational field is able to induce the tidal disruption of the star. If the tidal disruption happens outside the ISCO, some debris are left outside the BH. In the opposite case, the NS directly plunges into the BH and no matter is left outside. The NS fate depends on the NS tidal deformability $\Lambda_\mathrm{NS}$ and thus on the equation of state (EoS) for nuclear matter, on the BH and NS mass ratio $q$ (defined as $m_{\rm BH}/m_{\rm NS}>1$), and on the BH spin $\chi_{\rm BH}$ \citep{Shibata11, Foucart2012,Kyutoku2015,Kawaguchi2015,Foucart2018}. As we discussed in \cite{Barbieri2019}, the NS is subject to partial disruption for larger values of $\Lambda_\mathrm{NS}$, larger BH spins and lower mass ratios \citep{Bildsten1992,Shibata2009,Foucart2013a,Foucart2013b,Kawaguchi2015,Pannerale2015a,Pannerale15b,Hinderer2016,Kumar2017}. 

If the NS is disrupted, some neutron-rich material is released. The tidal debris comprise a gravitationally bound component, which forms an accretion disc around the final BH, and an unbound one, the so-called ``dynamical ejecta'' \citep{DiMatteo2002,Chen2007,Shibata11,Foucart2012,Janiuk2013,Kawaguchi2015}. When this matter is present outside the BH,  EM emission is expected to emerge from a variety of processes.  Therefore, beyond the study of the GW signal, further information on the binary can be extracted by detecting and characterizing the EM counterparts of such merger. 

The first GW signal from a NSNS merger was detected in 2017 \citep{GW170817} and an EM counterpart was soon observed \citep{gw170817em}, giving birth to the GW-EM multi-messenger astronomy. This observation of GW170817 demonstrated that NSNS mergers can be the progenitors of short duration gamma-ray bursts (SGRBs) \citep[as proposed by, e.g., ][]{Eichler1989,Narayan1992} and that they lead to the production of $r$-process elements whose decay powers the emission from the expanding ejecta, which produces an ultraviolet-optical-infrared transient called ``kilonova''  \citep{Lattimer1974,Li1998,Metzger2017}.

From the GW signal analysis, constraints on the binary combined tidal deformability were obtained, with an upper limit $\tilde{\Lambda}\leq800$ \citep{GW170817,GW170817_2}. This quantity is a mass-weighted combination of the two NS dimensionless tidal deformability parameters \citep{Raithel2018}. For which concerns the individual tidal deformabilities $\Lambda_1$ and $\Lambda_2$, assuming a common EoS, \cite{De2018} found - at
90\% confidence level - 
$\Lambda_1\lesssim500$ and $\Lambda_2\lesssim1200$ (assuming uniform mass prior), $\Lambda_1\lesssim700$ and $\Lambda_2\lesssim750$ (assuming mass prior from known NSNS systems),  $\Lambda_1 \lesssim600$ and $\Lambda_2\lesssim900$ (assuming mass prior from Galactic NSs). \cite{Radice2018_3} found constraints on $\tilde{\Lambda}$ from GW170817 multimessenger observations. Indeed, beside the cited upper limit coming from GW signal analysis, they found a lower limit $\tilde{\Lambda}\geq400$, necessary in order to produce enough ejecta to power a kilonova consistent with the observed one.

We note that while for a NSNS binary $\tilde{\Lambda}$ is a mass-weighted quantity of the two component deformabilities, for a BHNS binary, being the deformability of a BH equal to zero, $\tilde{\Lambda}$ is linked to the NS deformability parameter $\Lambda_\mathrm{NS}$ (through a coefficient containing the BH and NS masses).

Also BHNS mergers can produce a kilonova. Furthermore, there are indications that a relativistic jet could be launched from the remnant BH - accretion torus system \citep{Blandford1977,Tchekhovskoy2010} in presence of strong magnetic field amplification through magneto-rotational and Kelvin-Helmholtz instabilities in the torus \citep{Price2006,Zrake2013,Giacomazzo2015}. GR magneto-hydrodynamical simulations are on the verge of resolving the jet launch \citep[see e.g.][]{Shapiro-jet2015,Shapiro2017,Pasch2017,Ruiz-Jet-2018}, which could then power a GRB and, upon interaction of the jet with the interstellar medium (ISM), its afterglow. Kilonova emission from BHNS mergers was studied through semi-analytical models and fitting formulae from general-relativistic numerical simulations \citep{Kawaguchi2016} and radiation-transfer simulations \citep{Tanaka13,Tanaka2014,Fernandez2017}.

In \cite{Barbieri2019} we developed a suite of semi-analytical models to predict the properties of the EM counterparts of BHNS mergers. We included the nuclear-decay-powered kilonova and its radio remnant, the GRB prompt emission and the afterglow from the relativistic jet, covering all the EM spectrum from radio to gamma-rays.  The main focus in that work was the role of the BH spin in determining the degree of mass loss following the partial disruption of the NS and in determining the light curves associated to both the bound and unbound material.
We found that light curves feature universal traits but at the same time they show a high degree of degeneracy with respect to the BH mass and spin combinations.

In this paper, for simplicity, we consider only the kilonova and the GRB afterglow emissions. We expand on our previous work by exploring the dependence of the light curves on the BH mass and spin \citep[as in][]{Barbieri2019} but this time considering also BHs in the lower ``mass gap''. If such light stellar BHs form, the low $q$ values favour BHNS mergers that leave behind a large amount of debris and can be associated to the brightest EM events, within this new class of GW sources. As discussed in \cite{Littenberg2015}, if BHs are allowed to have small masses, close to the maximum NS mass, then for a binary merger with a total mass $\approx2 M_\mathrm{NS,max}$, it would be difficult to disentangle the nature of the binary analysing the GW signal alone, since the detector sensitivity to the final part of the inspiral and to the merger is hampered by high-frequency shot noise. Unless the SNR is very high, it can be difficult to discriminate a low mass BHBH system from a high mass NSNS system or a BHNS system. Combined GW and EM observations could help to disentangle these possibilities, and for this reason we investigate in detail the joint GW and EM detectability. 

We also investigate on how the EM counterparts to BHNS coalescences depend on the tidal deformability of the NS, and in turn on the EoS.

To this purpose we explore an interval of NS masses and a set of EoS. The BH mass and spin are fixed to selected values, to let the NS be partially disrupted prior to the final plunge.  We also take the opportunity, being this paper published in a special issue, to detail the semi-analytical model used in \cite{Barbieri2019} for sake of completeness and clarity.


The paper is organized as follows. In \S\ref{sec:bh_properties} we present and justify our choice of BH parameters. In \S\ref{sec:ejecta} we show the dependence of the ejecta mass on BH properties (for a given NS parameter set) and on NS properties (fixing the BH parameters). In \S\ref{sec:kilonova}
we describe in detail our model for kilonova emission. In \S\ref{sec:grb_afterglow} we present our model for relativistic jet launch and our assumptions on its structure, and we give the details for the GRB afterglow emission model. In \S\ref{sec:kn+grb} we show some example light curves in order to stress their dependence on NS properties. Finally, in \S\ref{sec:energy_kn} we show how the total energy radiated in the kilonova depends on BH and NS properties.


\section{Black hole parameters}\label{sec:bh_properties}

In this work, we select two BH masses, $6\msun$ and $3\msun$. The mass of $M_\mathrm{BH}=6\msun$ is taken as reference for the rapid SN explosion. This value is compatible with the mass distribution for stellar-mass BHs presented in \cite{Farr2011}. This is also a typical BH mass value from BHNS population synthesis models \citep{Dominik2012,Giacobbo2018}. The mass of $M_\mathrm{BH}=3\msun$ is taken instead as reference for the delayed SN explosion scenario, and is close to the maximum mass of a NS predicted from the most extreme NS EoS, which have been excluded by recent observations \citep{Radice2018_3}. Heavy stellar BHs with masses in excess of $10\msun$ are not considered here as in that case the NS is not tidally disrupted in general.

For the reasons explained above, we select two values for the BH spin, namely $\chi_{\rm BH}=0.5$ and $0.8$. In \S~\ref{sec:kn+grb} we also discuss how a BH spin $\chi_\mathrm{BH}=0.3$ would affect the light curves.

We assumed the BH spin to be aligned with the binary angular momentum, thus we fixed the tilt angle $\iota_\mathrm{tilt}=0$. In the following analysis we will show results for the four combinations of these BH parameters.

\section{Ejecta from BHNS mergers and the role of tidal deformability}\label{sec:ejecta}

The BH is described by its mass $M_{\rm BH}$ and dimensionless spin parameter $\chi_{\rm BH}$, which determine the radius of the innermost stable circular orbit, $R_{\rm ISCO}$. As the NS gets closer to the BH, tidal forces increase. When the relative distance reaches $d_{\rm tidal}\sim (M_\mathrm{BH}/M_\mathrm{NS})^{1/3}R_\mathrm{NS}$, the tidal gravity gradient over the stellar radius $R_\mathrm{NS}$ induced by the BH equals the NS self-gravity. Below this distance the NS undergoes partial disruption, spreading neutron-rich material in its surroundings.

If $d_{\rm tidal}<R_{\rm ISCO}$, the tidal disruption occurs too close to the BH and the NS plunges entirely into the BH, leaving no mass outside: in this case, we do not expect any EM counterpart. If instead $d_{\rm tidal}>R_{\rm ISCO}$, the NS is partially disrupted and the remnant BH is surrounded by neutron-rich matter. In this case it is possible to produce the EM counterparts.

The total mass $M_\mathrm{out}$ of matter left outside the BH can be divided into two components: the accretion disc, representing the gravitationally bound material, and the dynamical ejecta, the unbound part. We denote their masses as $M_\mathrm{disc}$ and $M_\mathrm{dyn}$, respectively.

For a fixed BH mass, a more massive NS requires a larger BH spin in order to be tidally disrupted. Indeed high-mass NSs are generally more compact, thus $d_\mathrm{tidal}$ is smaller and also $R_{\rm ISCO}$ must be small for the BH gravity gradient to unbind material. This translates into a larger required BH spin. For a fixed NS mass, more massive BHs have larger $R_\mathrm{ISCO}$, unless their spin is very high. Thus also in this case a high BH spin is required to avoid a direct plunge. Therefore, the best combination of binary parameters, necessary to induce the NS tidal disruption and release neutron-rich matter, requires low mass ratios $q=M_\mathrm{BH}/M_\mathrm{NS}$ and high BH spins.
 
At fixed BH and NS masses, the release of ejecta depends on the NS EoS or, equivalently, on its tidal deformability
defines as

\begin{equation}
\Lambda_\textrm{NS}=\frac{2}{3}k_2C_{\rm NS}^{-5},
\end{equation}
where $C_{\rm NS}=GM_{\rm NS}/(R_{\rm NS}c^2)$ is the NS compactness equal to  (with $c$  and $G$ is the speed of light and gravitational constant), and 
$k_2$ is the dimensionless tidal Love number defined as  $k_2=(3/2)G\lambda R_{\rm NS}^{-5}$ \citep{Flanagan2008}, with $\lambda$ the quadrupolar polarisability, 
representing the ratio of the induced quadrupole moment $Q_{ij}$ to the applied tidal field $E_{ij}$ ($Q_{ij}=-\lambda E_{ij}$).

\begin{figure}
    \centering
    \includegraphics[]{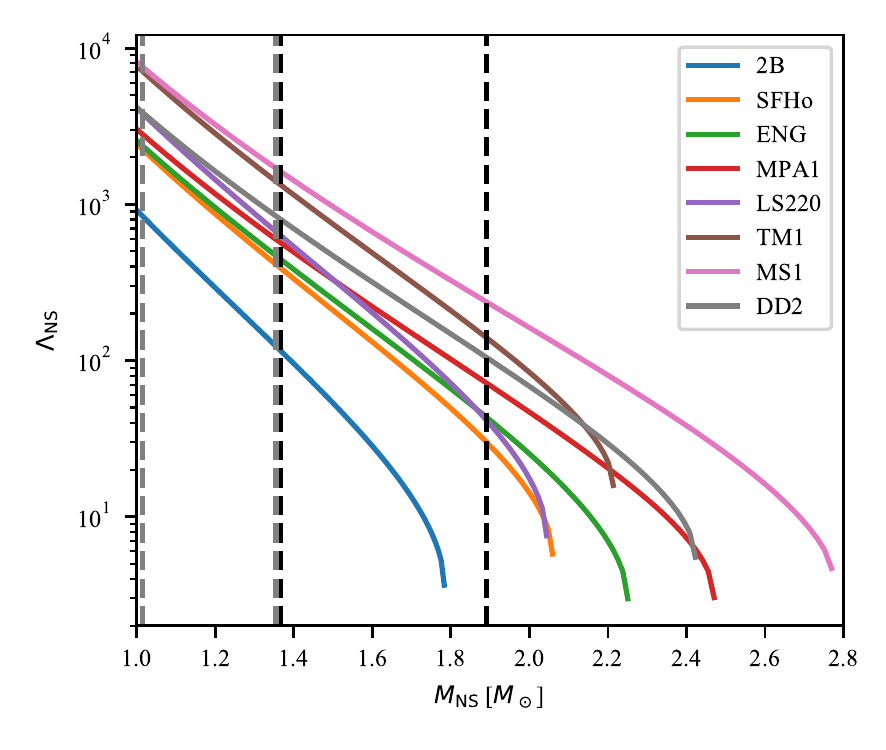}
    \caption{NS dimensionless tidal deformability parameter $\Lambda_\mathrm{NS}$ as a function of the NS mass $M_\mathrm{NS}$ for a set of selected EoS. The softest considered EoS is 2B (blue), while the stiffest one is MS1 (pink). Vertical dashed lines represent the estimated ranges for NS masses coming from GW170817 analysis (black for the primary component, gray for the secondary).}
    \label{fig:lambda-m}
\end{figure}

 For a given $M_\mathrm{NS}$, ``soft'' EoS lead to NSs with a smaller radius, a higher compactness $C_\textrm{NS}$ and consequently a smaller $\Lambda_\mathrm{NS}$ compared to ``stiff'' EoS. Thus, soft EoS conspire against tidal disruption.  Fig.~\ref{fig:lambda-m} shows the relation between $\Lambda_\mathrm{NS}$ and $M_\mathrm{NS}$ for a set of EoS, ranging from the softest 2B to the stiffest MS1. In this figure we also indicate the estimated mass ranges for the two NSs in the GW170817 event \citep{GW170817_2}.
Therefore, for small values of $\Lambda_\textrm{NS}$ (i.e.~soft EoS) the BH spin required to produce a considerable amount of ejecta, $M_\mathrm{out}$, needs to be large.

As in \cite{Salafia2017} for NSNS binaries, we parametrise the mass in the disc and ejecta  as functions of the BH and NS intrinsic parameters using fitting formulae from numerical relativity simulations.
We estimate $M_\mathrm{out}$ using the formula from \cite{Foucart2018}, whose free parameters are calibrated on numerical simulations of BHNS mergers. $M_\mathrm{out}$ depends on $M_\mathrm{BH}$, $M_\mathrm{NS}$, $M^\mathrm{b}_\mathrm{NS}$, $\chi_\mathrm{BH}$ and $\Lambda_\textrm{NS}$

\begin{equation}
M_\mathrm{out} = M^\mathrm{b}_\mathrm{NS}\left[\mathrm{max}\left(\alpha\frac{1-2\rho}{\eta^{1/3}}-\beta\tilde{R}_\mathrm{ISCO}\frac{\rho}{\eta}+\gamma,0\right)\right]^\delta,
\label{eq:mass-out}
\end{equation}
where $M^\mathrm{b}_\mathrm{NS}$ is the NS baryonic mass, $\eta=q/(1+q)^2$ is the symmetric mass ratio, $\rho=(15\Lambda_\mathrm{NS})^{-1/5}$, $\tilde{R}_\mathrm{ISCO}=R_\mathrm{ISCO}c^2/GM_\mathrm{BH}$ is the dimensionless ISCO.  The parameter in  Eq.~ \ref{eq:mass-out}  are those in \cite{Foucart2018}.

We calculate $M_\mathrm{dyn}$ and the ejecta rms velocity $\mathrm{v}_\mathrm{dyn}$ using the formula from \cite{Kawaguchi2016}. Here $M_\mathrm{dyn}$ depends on $M_\mathrm{BH}$, $M_\mathrm{NS}$, $M^\mathrm{b}_\mathrm{NS}$, $\chi_\mathrm{BH}$, $C_\textrm{NS}$ and $\iota_\mathrm{tilt}$, the angle between the BH spin and the binary total angular momentum:  
\begin{equation}\label{eq:Mdyn}\begin{split}
M_\mathrm{dyn}=M^\mathrm{b}_\mathrm{NS}&\Big\lbrace\mathrm{max}\left[ a_1q^{n_1}(1-2C_\mathrm{NS})/C_\mathrm{NS} - a_2q^{n_2}\tilde{R}_\mathrm{ISCO}(\chi_\mathrm{eff}) +\right.\\&\left.+ a_3(1-M_\mathrm{NS}/M^\mathrm{b}_\mathrm{NS})+a_4,0\right] \Big\rbrace,\end{split}
\end{equation}
where $\chi_\mathrm{eff}=\chi_\mathrm{BH}\cos{\iota_\mathrm{tilt}}$ is the effective BH spin, and the parameter values are those in \cite{Kawaguchi2016}. The rms velocity of the ejecta  depends only on $q$, and reads 
\begin{equation}
\mathrm{v}_\mathrm{dyn}=(aq+b)c,
\end{equation}
with  $a,b$ and $c$ here given in \cite{Kawaguchi2016}.

For the BHNS binary configurations explored in the simulations (focused on a BH mass range $\sim$ 4 -- 10$\msun$), the mass in the unbound ejecta  never exceeds few percent of the NS mass. We assume the limit for the dynamical ejecta mass as
\begin{equation}
    M_\mathrm{dyn,max}=f~M_\mathrm{out}.
\end{equation}
The factor $f$ represents the maximum ratio between the dynamical ejecta mass and the total mass remaining outside the BH. $f$ cannot exceed $0.5$. This value corresponds to results from numerical simulations of tidal disruption events (TDE) of an
unbound star orbiting a massive BH, where half of the star's mass is unbounded and the other half forms an accretion disc \citep{Rees1988}. However a NS is bound to the BH prior merging and  tidal disruption will proceed differently. \cite{Foucart2019} performed new simulations of BHNS mergers in the near-equal-mass regime, considering BH of mass near $1.4\msun$. The configuration producing the largest unbound component has $\sim~28$\% of the mass remaining outside the BH in dynamical ejecta. Therefore we assume $f=0.3$.


In order to calculate $M_\mathrm{dyn}$ we need to correlate the tidal deformability parameter with the stellar compactness. \cite{Yagi2017} derived an EoS-quasi-independent relation between $C_\mathrm{NS}$ and $\Lambda_\mathrm{NS}$ known as the ``C-Love'' relation: 
\begin{equation}\label{eq:Cns}
C_{\rm NS}=\sum_{k=0}^2a_k({\rm{ln}}\Lambda_{\rm NS})^k.
\end{equation}
We calculate $C_\mathrm{NS}$ using this formula with the coefficients $a_k$ as given in their work.

The NS baryonic mass in Eq.~ \ref{eq:Mdyn} is given by the relation
\begin{equation}\label{eq:Mb}
M_\textrm{NS}= M^\textrm{b}_{\rm NS} - B.E. \end{equation}
where $B.E.$ is the binding energy when $N$ baryons are assembled. \citep{Lattimer2001} related this binding energy to $M_\mathrm{NS}$ and $C_\mathrm{NS}$ to yield
\begin{equation}
B.E. = M_\textrm{NS}\frac{0.6C_\textrm{NS}}{1-0.5C_\textrm{NS}}.
\end{equation}
Thus we obtain
\begin{equation}
M^\textrm{b}_{\rm NS} = M_\textrm{NS}\left(1+\frac{0.6C_\textrm{NS}}{1-0.5C_\textrm{NS}}\right),
\end{equation}

\noindent
With these definitions we are able to compute the total mass remaining outside the BH and the dynamical ejecta mass. Therefore the mass in the disc is simply their difference:
\begin{equation}
M_{\rm disc}=\max\left[M_{\rm out}-M_{\rm dyn},0\right].
\end{equation}

\begin{figure}
    \centering
    \includegraphics[]{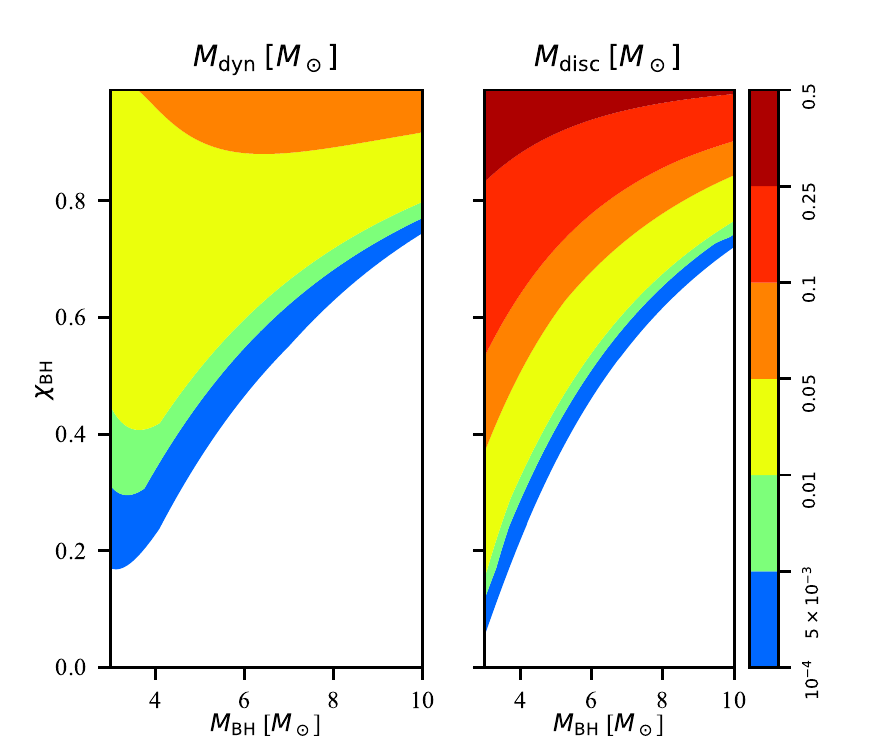}
    \caption{Dynamical ejecta (left) and disc masses (right) produced during the merger in the $M_\mathrm{BH}-\chi_\mathrm{BH}$ parameter space, assuming a NS with $M_\mathrm{NS}=1.4~\msun$ and $\Lambda_\mathrm{NS}=330$ (corresponding to SFHo EoS).}
    \label{fig:ejecta_BH}
\end{figure}

In this work we consider BHs with masses $3~M_\odot<M_\mathrm{BH}<5~M_\odot$, while in \cite{Barbieri2019} we limited our analysis to $M_\mathrm{BH}>6~M_\odot$ (assuming the rapid SN explosion). Therefore, before analysing the dependence of the mass in the ejecta on NS properties, we extended our previous analysis to lower BH masses. We fixed the NS properties, namely $M_\mathrm{NS}=1.4~M_\odot$ and $\Lambda_\mathrm{NS}=330$, corresponding to the SFHo EoS \citep{SFHo}. In Fig.~\ref{fig:ejecta_BH} we show how the BH mass and spin affect the production of dynamical ejecta and disc masses. White regions represent parameter combinations corresponding to a direct plunge of the NS, thus leaving no mass outside the BH, and thus no EM counterpart. It is clear that, for a given pair of $M_\mathrm{NS}$ and $\Lambda_\mathrm{NS}$, larger dynamical ejecta and disc masses are associated to less massive and/or faster spinning BHs.

\begin{figure}
    \centering
    \includegraphics[width=\columnwidth]{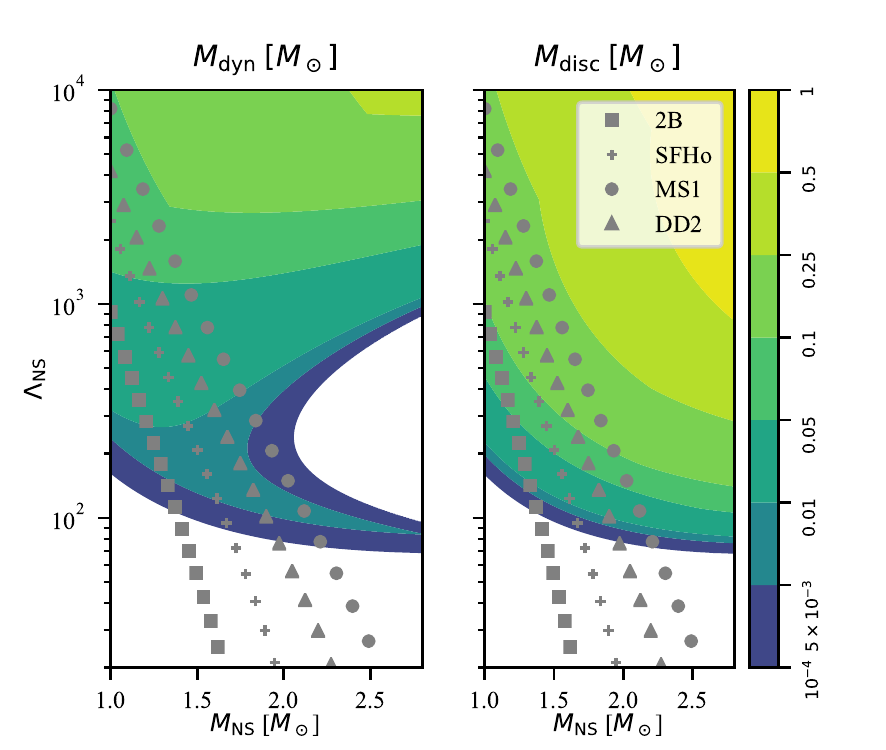}
    \caption{Dynamical ejecta (left) and disc masses (right) produced during the merger in the $M_\mathrm{NS}-\Lambda_\mathrm{NS}$ parameter space, assuming a BH with $M_\mathrm{BH}=3\msun$ and $\chi_\mathrm{BH}=0.5$. The NS compactness is computed assuming the C-Love relation. Gray symbols show the $\Lambda_\mathrm{NS}-M_\mathrm{NS}$ relation for a set of EoS. The match of any point with the underlying shaded area corresponds to the estimated value of the dynamical ejecta and disc masses.}
    \label{fig:ejecta_m3_s5}
\end{figure}

\begin{figure}
    \centering
    \includegraphics[width=\columnwidth]{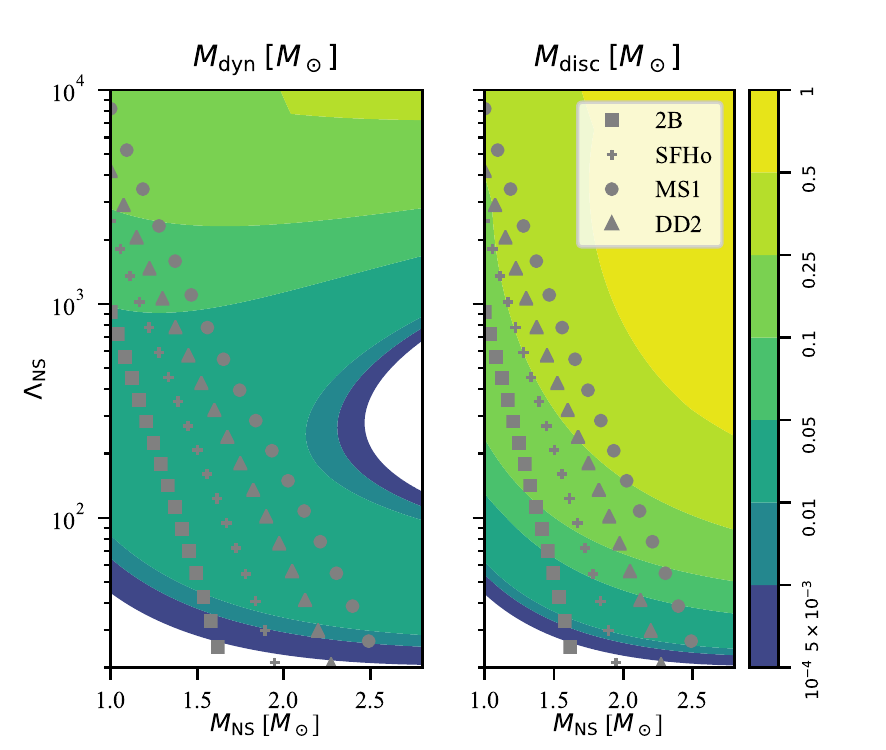}
    \caption{Same as Fig. \ref{fig:ejecta_m3_s5}, assuming a BH with $M_\mathrm{BH}=3\msun$ and $\chi_\mathrm{BH}=0.8$.}
    \label{fig:ejecta_m3_s8}
\end{figure}

\begin{figure}
    \centering
    \includegraphics[width=\columnwidth]{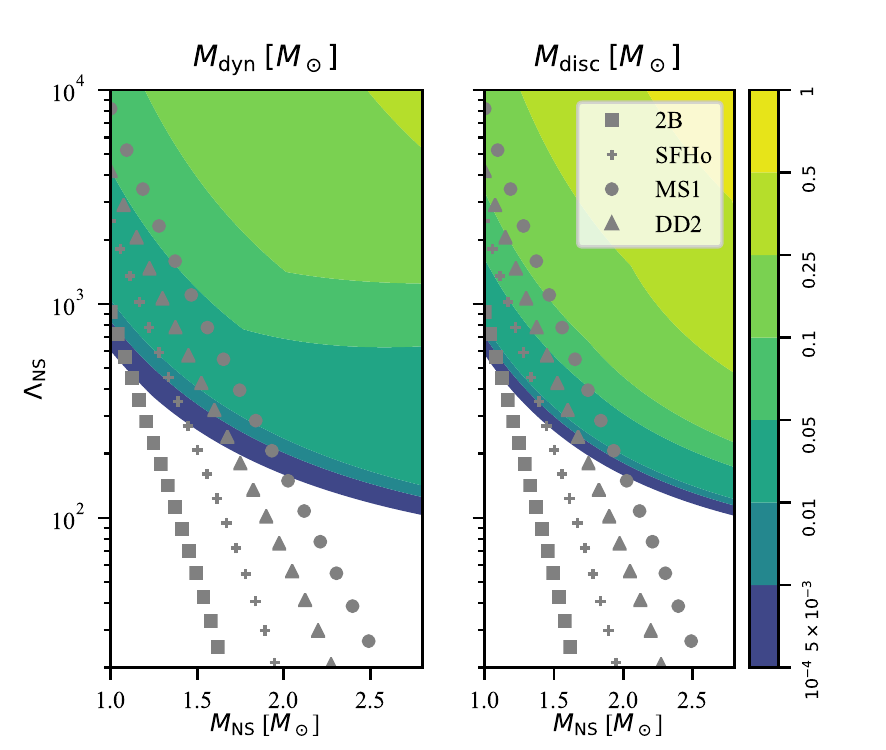}
    \caption{Same as Fig. \ref{fig:ejecta_m3_s5}, assuming a BH with $M_\mathrm{BH}=6\msun$ and $\chi_\mathrm{BH}=0.5$.}
    \label{fig:ejecta_m6_s5}
\end{figure}

\begin{figure}
    \centering
    \includegraphics[width=\columnwidth]{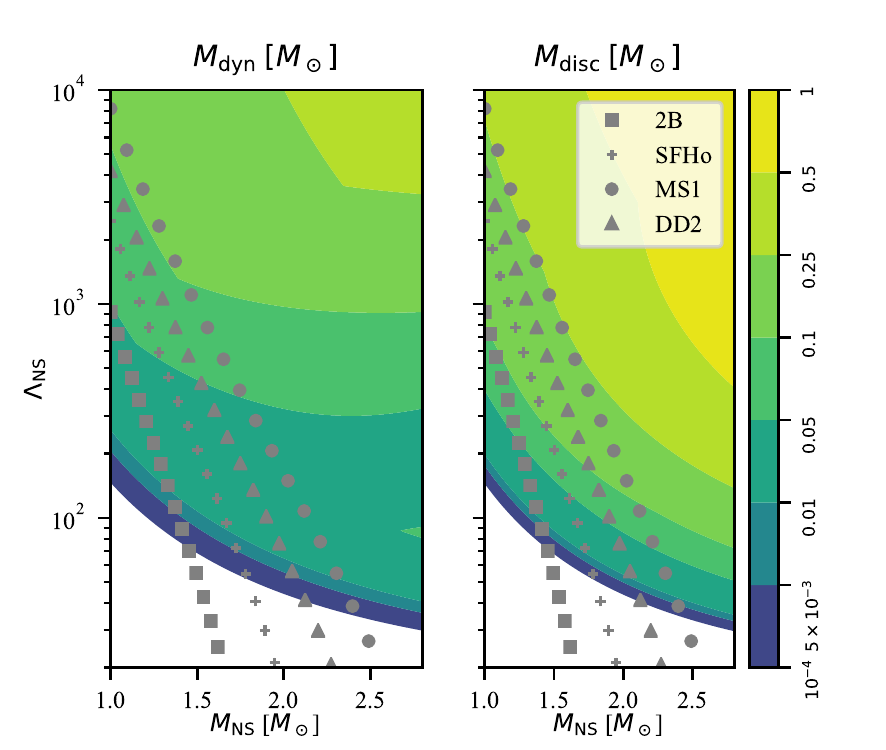}
    \caption{Same as Fig. \ref{fig:ejecta_m3_s5}, assuming a BH with $M_\mathrm{BH}=6\msun$ and $\chi_\mathrm{BH}=0.8$.}
    \label{fig:ejecta_m6_s8}
\end{figure}

Figure \ref{fig:ejecta_m3_s5} shows the dynamical ejecta and disc masses produced during the merger in the $M_\mathrm{NS}-\Lambda_\mathrm{NS}$ plane. As an indication we plot with gray symbols the $\Lambda_\mathrm{NS}-M_\mathrm{NS}$ relation for some EoS. It is evident that in principle the optimal condition to produce massive dynamical ejecta and discs fixing $M_\mathrm{BH}$ and $\chi_\mathrm{BH}$ is having a massive NS and/or a large $\Lambda_\mathrm{NS}$, if each NS parameter pair was possible. However only certain $M_\mathrm{NS}-\Lambda_\mathrm{NS}$ pairs are described by a physically motivated EoS. Limiting to existing EoS, we find that the optimal condition to produce massive dynamical ejecta and discs fixing $M_\mathrm{BH}$ and $\chi_\mathrm{BH}$ is having a low mass NS. Indeed, as shown in Fig.~\ref{fig:lambda-m}, for each EoS low mass NSs correspond to the largest values of tidal deformability.
Figures \ref{fig:ejecta_m3_s8}-\ref{fig:ejecta_m6_s5}-\ref{fig:ejecta_m6_s8} show the same for different BH parameters. We see that, when fixing the NS properties, the optimal condition to produce massive dynamical ejecta and discs is having a low mass BH with a large spin.

It is interesting to note that NSs never suffer a total tidal disruption in the considered cases. Indeed in all the $\Lambda_\mathrm{NS}-M_\mathrm{NS}$ parameter space the total mass remaining outside the BH is always $\lesssim42\%~M_\mathrm{NS}$. This upper limit is reached for extreme combinations of NS parameters (in particular very large $\Lambda_\mathrm{NS}$), that are not described by any physically motivated EoS. If we limit the analysis to MS1 EoS (gray dots), the stiffest one among those presented, we find $M_\mathrm{out}\lesssim40\%~M_\mathrm{NS}$. Gray crosses represent the SFHo EoS, which we consider likely more realistic. This EoS is compatible with nuclear and astrophysical constraints and it associates to a $1.4\msun$ NS a radius of $\sim12$ km, compatible with the NS radii estimates in the GW170817 signal analysis \citep{LVC_EoS2017}. Considering the SFHo EoS, we find that $M_\mathrm{out}\leq32\%~M_\mathrm{NS}$.  

We stress that fits from \cite{Kawaguchi2016} are based on simulations with $3\leq q \leq7$ and $300\leq\Lambda_\mathrm{NS}\leq1500$, while fits from \cite{Foucart2018} on simulations in the ranges $1\leq q\leq7$ and $280\leq \Lambda_\mathrm{NS}\leq2070$. Therefore dynamical ejecta estimates for the case with $M_\mathrm{BH}=3$ and, consequently, $q<3$ are extrapolations from the fit, such as points with $\Lambda_\mathrm{NS}$ above or below the indicated ranges. We thus consider our results outside these ranges as only indicative. \cite{Foucart2019} compared their results from simulations of near-equal-mass BHNS mergers ($1<q<1.9$) with fits from \cite{Kawaguchi2016}, finding that the dynamical ejecta mass fit is quite consistent with simulations also in this low mass ratio regime, while extrapolation of the dynamical ejecta velocity fit overestimates by a factor $\sim$2 their results.


\section{Kilonova model}\label{sec:kilonova}

BHNS mergers are possible sites for heavy elements production, through $r$-process nucleosynthesis \citep{Lattimer1974}. This process happens in the neutron-rich ejecta released during the NS tidal disruption with synthesized nuclei far from the valley of stability. These nuclei undergo radioactive decay, powering the kilonova (KN) emission on timescales ranging from $\sim$hours to $\sim$weeks after the merger.

Different matter ejection mechanisms can be identified, each associated to a typical timescale and ejecta properties:
\begin{enumerate}
\item {Dynamical ejecta}, produced on a short timescale ($\sim$ms) by tidal interactions \citep[e.g.][]{Kawaguchi2016,Radice2018_2};
\item {Wind ejecta}, produced on a timescale of tens of milliseconds by the accretion disc during the initial neutrino-cooled phase \citep{Just2015} through neutrino-matter interactions and magnetic pressure \citep[e.g.][] {Dessart2009,Kiuchi2015,Fernandez2017};
\item{Viscous ejecta}, produced through magnetically originated viscous processes inside the disc during the advection-dominated phase \citep{Just2015}. The timescale of this ejection is comparable with the accretion duration, being related to the angular momentum transport, viscous heating and nuclear recombination \citep[e.g.][] {Fernandez2013,Radice2018}.
\end{enumerate}

\subsection{Wind and secular ejecta emission}
We followed \cite{Perego2017} to calculate the kilonova
emission from wind and secular ejecta. Their semi-analytical model assumes symmetry with respect to the system total angular momentum direction and divides the polar angle $\theta$ into 30 equally spaced (in $\cos{\theta}$) slices. The parameters for each component are the mass $m_\mathrm{ej}$ and its angular distribution, the average velocity $\mathrm{v}_\mathrm{ej}$ in the radial direction and the effective grey opacity $\kappa_\mathrm{ej}$, possibly depending on $\theta$. They assume that, inside each slice, matter expands homologously. Numerical simulations \citep[e.g.~][]{Rosswog2013} and analytical arguments \citep{Wollaeger2018} show that the matter distribution in velocity space can be broadly described as 
\begin{equation}\label{eq:m_v_dist}
dm/d\mathrm{v} \propto (1- \left( \mathrm{v}/\mathrm{v}_{\rm max} \right)^2)^3,
\end{equation}
with $\mathrm{v}_{\rm max}$ the largest velocity of ejecta. In the following we use the velocity $\mathrm{v}$ as a Lagrangian coordinate. We refer to the ejecta part moving at a certain velocity as a ``shell''. The relation between maximum and rms mean velocity is $\mathrm{v}_{\rm max} = 128/35 \, \mathrm{v}_{\rm ej}$. For each angular slice, thermal emission from the photospheric radius is calculated following \cite{Grossman2014} and \cite{Martin2015}. 

For each component $i$, the bolometric luminosity emitted at time $t$ in the angular slice $j$ is 
\begin{equation}\label{eq:kn_lum}
L^\mathrm{bol}_{ij}(t,m_{ij},Y_{\mathrm{e},ij})=\dot{\epsilon}_\mathrm{nuc}(t,Y_{\mathrm{e},ij})~m_{\mathrm{rad},ij}(t,m_{ij},Y_{\mathrm{e},ij}).
\end{equation}
In this equation $m_{ij}$ and $Y_{\mathrm{e},ij}$ are the mass and electron fraction of the $i$-th component in the $j$-th angular slice. Since the treatment is identical, in what follows we focus on a single angular slice and a single component $ij$ and we drop the indices for clarity. Based on detailed nucleosynthesis calculations, \cite{Korobkin2012} obtained a simple analytical fitting formula for the nuclear heating rate $\dot{\epsilon}_\mathrm{nuc}$, namely
\begin{equation}
 \dot{\epsilon}_\mathrm{nuc}(t)=\epsilon_0~\frac{\epsilon_{th}}{0.5}~\left[\frac{1}{2}-\frac{1}{\pi}\arctan\left(\frac{t-t_0}{\sigma}\right)\right],
\end{equation}
where $\epsilon_0\approx10^{18-19}$~erg~s$^{-1}$~g$^{-1}$, $\epsilon_\mathrm{th}$ is the thermalization efficiency, $\sigma=0.11$~s and $t_0=1.3$~s. \cite{Perego2017} introduced an electron-fraction-dependent term $\epsilon_{Y_e}$ in $\dot{\epsilon}_\mathrm{nuc}$ in order to take into account that ejecta with large electron fraction (and therefore low opacity) have a decay half-life of few hours, thus their emission is "boosted" at early times:
\begin{equation}
\epsilon_{Y_e}(t)=\begin{cases}0.5+2.5\left\lbrace 1+\exp\left [4\left (t/1\mathrm{d}-1\right)\right]\right\rbrace^{-1}& \mbox{if $Y_e\geq0.25$} \\ 1& \mbox{otherwise.}
\end{cases}\end{equation}
Here $m_{\mathrm{rad},ij}$ is the mass of the radiating shell for the $i$-th component in the $j$-th angular slice. The radiating shell is the part of the ejecta comprised between the diffusion surface (where the optical depth $\tau=\kappa\bar{\rho}\Delta r=c/\mathrm{v}$, below which the diffusion timescale is larger than the dynamical one) and the photosphere (where $\tau=2/3$, above which the average density is too low and photon thermalisation is not efficient). For homologous expansion, the mass distribution in velocity space is stationary. We can define the mass moving faster than a certain velocity $\mathrm {v}$ as:
\begin{equation}\label{eq:m_v}
m_{>\mathrm{v}}(\mathrm{v})~=~m_\mathrm{ej}~\left[1+F\left(\frac{\mathrm{v}}{\mathrm{v}_\mathrm{max}}\right)\right],     
\end{equation}
where $F(x)$, for the mass distribution given in Eq.~\ref{eq:m_v_dist},  is
\begin{equation}
F(x)=\frac{35}{112}x^7-\frac{105}{80}x^5+\frac{35}{16}x^3-\frac{35}{16}x.    
\end{equation}
We consider diffusion to become effective when the optical depth falls below $c/v$,
\begin{equation}
\tau=\kappa\bar{\rho}\Delta r=\frac{\kappa m_{>\mathrm{v}}}{4\pi(\mathrm{v}t)^2}=\frac{c}{\mathrm{v}},  
\end{equation}
thus the time at which the diffusion surface corresponds to the shell with velocity $v$ is given by
\begin{equation}
t_\mathrm{diff}=\sqrt{\frac{\kappa m_{>\mathrm{v}}}{4\pi \mathrm{v}c}}.    
\end{equation}
The time at which the photosphere corresponds to the shell with velocity $\mathrm{v}$ is
\begin{equation}
t_\mathrm{phot}=\sqrt{\frac{3km_{>\mathrm{v}}}{8\pi \mathrm{v}^2}}.    
\end{equation}
Inverting these relations we can find the time evolution of the diffusion and photospheric shells. We calculate the effective radiating mass as
\begin{equation}
m_\mathrm{rad}=m_{>\mathrm{v_\mathrm{diff}}}-m_{>\mathrm{v_\mathrm{phot}}}    
\end{equation}
The emission is assumed to be described by a blackbody. \cite{Barnes2013} noted that when the ejecta temperature reaches the first ionisation temperature of lanthanides $T_\mathrm{La}$, these elements recombine and there is a sharp drop in opacity. The photosphere follows the recombination front and recedes inward, maintaining a constant temperature, equal to the recombination value. The photospheric radius before recombination is simply
\begin{equation}
R_\mathrm{phot}=\mathrm{v}_\mathrm{phot}t,    
\end{equation}
while after recombination, using the relation $L_\mathrm{bol}=A\sigma_\mathrm{SB}T^4=R^2\Omega\sigma_\mathrm{SB}T^4$ we find the photospheric radius giving a constant temperature
\begin{equation}
R_\mathrm{phot}=\sqrt{\frac{L_\mathrm{bol}}{\Omega\sigma_\mathrm{SB}T_\mathrm{La}^4}},   
\end{equation}
where $\sigma_\mathrm{SB}$ is the Stefan-Boltzmann constant and $\Omega$ is the subtended solid angle. Therefore at each time the photospheric radius is the minimum of the two expressions. We can summarize the two cases in the following way:
\begin{equation}
\begin{cases} R_\mathrm{phot}=\mathrm{v}_\mathrm{phot}t,~T=\left(\frac{L_\mathrm{bol}}{\Omega\sigma_\mathrm{SB}R_\mathrm{phot}^2}\right)^{1/4} & \mbox{before recombination} \\ T=T_\mathrm{La},~R_\mathrm{phot}=\sqrt{\frac{L_\mathrm{bol}}{\Omega\sigma_\mathrm{SB}T_\mathrm{La}^4}} & \mbox{after recombination}\end{cases}    
\end{equation}
Following \cite{Martin2015} we compute the observed spectral flux by superposing Planckian distributions, projecting the emitting surface in each angular bin along the line of sight. Given the blackbody spectrum in each angular bin and the observer direction $\mathbf{w}$, the total observed spectral flux is
\begin{equation}
F_\nu(\mathbf{w},t)=\sum_j B_\nu(T_j(t))\int_{\hat{\mathbf{n}}_\textbf{j}\cdot\mathbf{w}>0}{\mathbf{w}\cdot d\mathbf{\Omega}},
\end{equation}
where $\hat{\textbf{n}}_\textbf{k}$ is the unit vector perpendicular to the photosphere in the $j$-th angular slice. The projection factors given by the integral are time independent, thus we compute them in advance and use them as weighting factors $p_j(\mathbf{w})$,
\begin{equation}
F_\nu(\mathbf{w},t)=\sum_j p_j(\mathbf{w}) B_\nu(T_j(t)).    
\end{equation}
At any given time $B_\nu$ is expressed as
\begin{equation}
B_\nu=\left(\frac{R_\mathrm{phot}}{d_\mathrm{L}}\right)^2\frac{2h\nu^3}{c^2}\frac{1}{e^{h\nu/(k_\mathrm{B}T)}-1},
\end{equation}
where $d_\mathrm{L}$ is the luminosity distance, $h$ the Planck constant, $\nu$ the considered frequency and $k_\mathrm{B}$ the Boltzmann constant.

\subsection{Dynamical ejecta emission}

\begin{figure}
    \centering
    \includegraphics[width=\columnwidth]{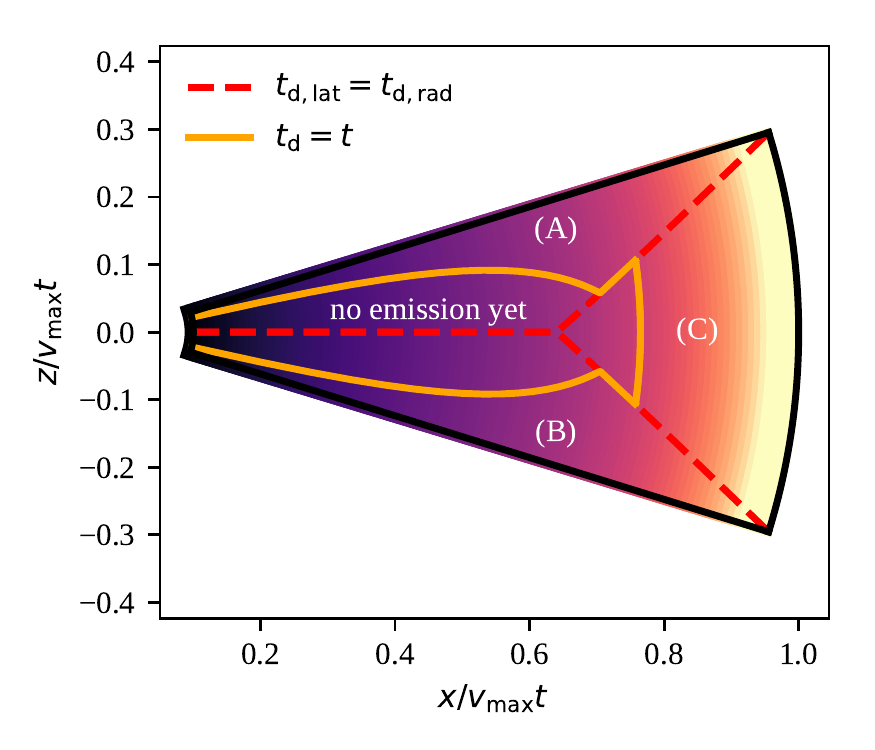}
    \caption{Schematic division of a section of dynamical ejecta (assumed to have a crescent-like geometry). The decreasing ejecta density in the outward direction is represented qualitatively by the colouring. The equatorial plane is at $z=0$. We identify three regions, delimited by the red-dashed line, according on the edge for which the diffusion time is the shortest. For region A it is the upper latitudinal edge, for region B the lower one, and for region C the radial edge. In ejecta regions above the orange solid line radiation can diffuse to the surface, while below that line emission is not possible because radiative diffusion has not yet reached the ejecta edge. Reproduced from \cite{Barbieri2019}.}
    \label{fig:dyn_ej_regions}
\end{figure}

In BHNS mergers the dynamical ejecta are not symmetric with respect to the rotational axis of the system. Indeed \cite{Kawaguchi2016} and \cite{Fernandez2017} showed that the typical geometry for this kind of ejecta is a crescent, lying close to the equatorial plane and extending in the azimuthal direction over approximately half of the plane ($\phi_\mathrm{dyn}\sim\pi$), and in the latitudinal direction over an angle $\theta_\mathrm{dyn}\approx0.2-0.5$ rad. \cite{Kawaguchi2016} gave an analytical model for the dynamical ejecta emission, assuming a uniform velocity distribution, based on rescaling of the spectrum from a radiative transfer simulation from \cite{Tanaka2014}. In \cite{Barbieri2019} we presented a simple semi-analytical model to treat this emission.

We assumed the dynamical ejecta mass to be distributed in velocity space as in Eq.~\ref{eq:m_v}. Each dynamical ejecta shell is assumed to emit from its latitudinal photosphere or towards the radial photosphere, based on which diffusion time is the shortest. We follow the same arguments on the diffusion approximation as above.
Defining $\theta$ as the angle measured from the equatorial plane, dynamical ejecta extend from $\theta=-\theta_\mathrm{dyn}$ to $\theta=\theta_\mathrm{dyn}$. We limit our discussion to the region above the equatorial plane($\theta>0$), as the same results hold for the region below, simply reversing the signs. The diffusion time in the latitudinal direction for photons diffusing upwards in the shell and produced at an angle $\theta$ can be expressed as
\begin{equation}
t_{\mathrm{d,lat}}\sim{\frac{(\theta_\mathrm{dyn}-\theta)^2\kappa_\mathrm{dyn}\, dm/d\mathrm{v}}{c\,\theta_\mathrm{dyn}\phi_\textrm{dyn}\,t}}
,\end{equation}
where $\kappa_\mathrm{dyn}$ is the opacity of dynamical ejecta. 
Instead the diffusion time in the radial direction is
\begin{equation}
t_{\mathrm{d,rad}}\sim{\frac{\kappa_\mathrm{dyn} m_{\mathrm{dyn},>\mathrm{v}}(\mathrm{v}_\mathrm{max}-\mathrm{v})}{c\,\theta_\mathrm{dyn}\phi_\mathrm{dyn} \mathrm{v}^2 t}}.
\end{equation}
We can find the angle $\theta_\mathrm{lat}(v)$ above which the diffusion time in the latitudinal direction is shorter than that in radial direction. It is
\begin{equation}
    \theta_\mathrm{lat}(\mathrm{v}) = \theta_\mathrm{dyn}- \min\left(\theta_\mathrm{dyn},\sqrt{\frac{m_{\mathrm{dyn},>\mathrm{v}}(\mathrm{v}_\mathrm{max}-\mathrm{v})}{\mathrm{v}^2 dm/d\mathrm{v}}}\right)
.\end{equation}
We use this angle to identify three regions in the ejecta, each one emitting only in the direction for which the diffusion time is the shortest, as illustrated in Fig.~\ref{fig:dyn_ej_regions}. We label A and B the regions emitting in the latitudinal direction above and below the equatorial plane, respectively, while we label C the region emitting in the radial direction. Inside region A, the diffusion time is equal to the elapsed time at an angle
\begin{equation}
    \theta_\mathrm{d}(\mathrm{v},t) = \theta_\mathrm{dyn}- t\sqrt{\frac{c\,\theta_\mathrm{dyn}\phi_\mathrm{dyn}}{\kappa_\mathrm{dyn}dm/d\mathrm{v}}}
.\end{equation}
Assuming a uniform density distribution in the latitudinal direction and that the nuclear heating energy release that happens above $\theta_\mathrm{d}$ instantaneously contributes to the latitudinal emission, the latitudinal luminosity per unit velocity can be expressed as
\begin{equation}\label{eq:lum_kn_lat}
   \frac{dL_\mathrm{lat}}{d\mathrm{v}}(\mathrm{v},t)= \frac{1}{2}\dot\epsilon(t) \frac{dm}{d\mathrm{v}} \times \max\left(1-\frac{\theta_\mathrm{lat}(\mathrm{v})}{\theta_\mathrm{dyn}},1-\frac{\theta_\mathrm{d}(\mathrm{v},t)}{\theta_\mathrm{dyn}}\right)
,\end{equation}
where the factor $1/2$ takes into account that we only considered the region above the equatorial plane.

For a given shell, the latitudinal surface is
\begin{equation}
\frac{dS_{\mathrm{lat}}}{d\mathrm{v}}(\mathrm{v},t)=\phi_\mathrm{dyn}\mathrm{v}\,d\mathrm{v}\,t^2
,\end{equation}
and the latitudinal annulus above the shell has an effective temperature given by
\begin{equation}
T_{\mathrm{BB,lat}}(\mathrm{v},t)=\left(\frac{dL_{\mathrm{lat}}/d\mathrm{v}}{\sigma_\mathrm{SB}(dS_{\mathrm{lat}}/d\mathrm{v})}\right)^{1/4}.
\end{equation}
Considering the effect of lanthanides recombination described above, we set
\begin{equation}\label{eq:Tbb}
T_{\mathrm{lat}}(\mathrm{v},t)=\max(T_{\mathrm{BB,lat}}(\mathrm{v},t),T_\mathrm{La}). 
\end{equation}

For the region emitting in radial direction (region C), the approach is similar, but we must take into account the relative speed between the shell and the emitting surface.  As before, the condition for which radiation can escape from a certain region is that the radial diffusion speed is larger than the local velocity. This happens beyond a `diffusion velocity' $\mathrm{v}_\mathrm{d}$, obtained through the implicit equation
\begin{equation}
    t = \sqrt{\frac{\kappa_\mathrm{dyn}m_{\mathrm{dyn},>\mathrm{v}_\mathrm{d}}}{\theta_\mathrm{dyn}\phi_\mathrm{dyn}\mathrm{v}_\mathrm{d}c}}
.\end{equation}

Therefore the radial luminosity is 
\begin{equation}\label{eq:lum_kn_rad}
L_\mathrm{rad}(t) = \dot \epsilon(t)~ m_{\mathrm{rad},>\mathrm{v}_\mathrm{d}}(t),
\end{equation}
where $m_{\mathrm{rad},>\mathrm{v}}$ is the mass moving faster than v in region C:
\begin{equation}
m_{\mathrm{rad},>\mathrm{v}}= \int_{\mathrm{v}}^{\mathrm{v}_\mathrm{phot}} \frac{\theta_\mathrm{lat}(\mathrm{v})}{\theta_\mathrm{dyn}} \frac{dm}{d\mathrm{v}}d\mathrm{v}.
\end{equation}
We perform this integral up to $\mathrm{v}_\mathrm{phot}$ in order to exclude from emission the material above the photosphere (as explained before).
The radial emission surface is
\begin{equation}
S_\mathrm{rad}(t)\sim\phi_\mathrm{dyn}\theta_\mathrm{dyn}\mathrm{v}_\mathrm{ph}^2 t^2 \, ,
\end{equation}
where the photospheric velocity is obtained by solving 
\begin{equation}
    \tau = \frac{2}{3} = \frac{\kappa_\mathrm{dyn} m_{>\mathrm{v}_\mathrm{ph}}}{ \theta_\mathrm{dyn}\phi_\mathrm{dyn}\mathrm{v}_\mathrm{ph}^2 t^2}\, .
\end{equation}

\noindent The radial photospheric effective temperature is
\begin{equation}
T_\mathrm{rad}(t)=\max\left[\left(\frac{L_{\mathrm{rad}}(t)}{\sigma_\mathrm{SB}S_\mathrm{rad}(t)}\right)^{1/4},T_\mathrm{La}\right]. 
\end{equation}
\\ Assuming that the dynamical ejecta is geometrically thin, for an observer whose line of sight forms an angle $\theta_\mathrm{view}$ with the polar axis the projection factor for latitudinal emission is
\begin{equation}
f_\mathrm{lat}=\cos(\theta_\mathrm{view}).    
\end{equation}
Instead for radial emission the projection factor is 
\begin{equation}\begin{split}
f_\mathrm{rad}=&\,\pi \cos(\theta_\mathrm{view})\sin^2(\theta_\mathrm{dyn})+\\&+2\sin(\theta_\mathrm{view})[\theta_\mathrm{dyn}+\sin(\theta_\mathrm{dyn})\cos(\theta_\mathrm{dyn})].    
\end{split}\end{equation}

\subsection{Kilonova light curves}\label{sec:klc}
Our final light curves are obtained by summing the fluxes from dynamical (both latitudinal and radial emission), wind and secular ejecta. 

Dynamical ejecta result directly from the NS tidal disruption, so they remain cold and are not significantly irradiated by neutrino emission from the inner disc. This preserves the low $Y_\mathrm{e}$ expected for NS matter. As shown in \cite{Roberts2017}, indeed, robust $r$-process nucleosynthesis always occurs inside this type of ejecta. Therefore, we associate a high opacity $\kappa_{\rm dyn} = 15 \,{\rm cm^2~g^{-1}}$ to this component. 


For the wind and secular ejecta, we assumed parameter values coming from the analysis of the kilonova associated with GW170817. However, to take into account the intrinsic properties of the different binary we are considering, we modified some of them. We calculate the wind and viscous ejecta masses as fractions $\xi_{\rm w}=0.01$ and $\xi_{\rm s}=0.2$ of the disc mass \citep{Just2015,Fernandez2013,Metzger2014}.
In the BHNS case, indeed, the wind ejecta is  notably a smaller fraction of the disc with respect to the NSNS case. This is due to the lack of a possible intermediate supra- or hypermassive NS state that would produce an intense neutrino wind. For the wind and secular ejecta we assume opacities $\kappa_\mathrm{w}=1 \,{\rm cm^2~g^{-1}}$ and $\kappa_\mathrm{s}=5 \,{\rm cm^2~g^{-1}}$ respectively.

We tested our model on the observed kilonova associated with GW170817, using typical ejecta parameters for the NSNS merger case. The light curves obtained with our model are consistent with observations (paper in preparation). In \S~\ref{sec:sel_eos} we compared inferred kilonova light curves with those from AT2017gfo, the kilonova associated with GW170817.

\section{GRB Afterglow model}\label{sec:grb_afterglow}

\subsection{Relativistic jet launch}

When the NS is tidally disrupted and some of the released material forms a disc, its accretion on the remnant BH can induce the launch of a relativistic jet through the Blandford-Znajek mechanism \citep{Blandford1977,Komissarov2001}. As shown by \cite{Tchekhovskoy2010}, this process can produce a luminosity
\begin{equation}
\label{eq:Lbz}
L_{\rm BZ}\propto \frac{G^2}{c^3} M_\mathrm{BH}^2 B^2 \Omega_{\rm H}^2 f(\Omega_\mathrm{H}),
\end{equation}
where $B$ is the amplitude of the magnetic field at the BH horizon, $0\leq \Omega_{\rm H}\leq 1/2$ is the dimensionless angular velocity evaluated at the horizon,
\begin{equation}
\Omega_{\mathrm {H}}=\frac{\chi_\mathrm{BH}}{2(1+\sqrt{1-\chi_\mathrm{BH}^2})},\end{equation}
and $f(\Omega_\mathrm{H}) = 1+1.38\Omega_\mathrm{H}^2-9.2\Omega_\mathrm{H}^4$ is a correction for high-spin values.\\
The mass and spin here refer to the remnant BH. We compute the spin using Eq.~11 from \cite{Pannarale2013}.

Assuming that, after the merger, the magnetic field $B$ is amplified by Kelvin-Helmholtz and magneto-rotational instabilities (MRI) up to a fixed fraction of the rest mass energy density of the disc \citep{Giacomazzo2015}, we can write
\begin{equation}
B^2 \propto \frac{c^5}{G^2} \dot{M}M_\mathrm{BH}^{-2},
\end{equation}
where $\dot{M}$ is the rate of mass accretion on the remnant BH. Therefore
\begin{equation}
\label{eq:Lbz_eff}
L_{\rm BZ}\propto \dot{M} c^2 \Omega_{\rm H}^2 f(\Omega_\mathrm{H}).
\end{equation}
This scaling is consistent with general-relativistic magneto- hydrodynamic (GRMHD) simulations of compact object mergers where a jet is launched \citep{Shapiro2017}. 

In principle, during its propagation, the jet may interact with the other ejecta loosing some energy. However, assuming that the jet is launched perpendicularly to the accretion disc, it likely encounters very low density ejecta. Indeed, in BHNS mergers the dynamical ejecta lie close to the equatorial plane \citep[from][$\theta_{\rm{dyn}}\leq 22^\circ$]{Kawaguchi2016}. This is due to the absence of shocks that would generate a more isotropic distribution of dynamical ejecta. This happens in NSNS mergers when the two stars collide, but in BHNS mergers the BH obviously does not have a ``crust''. Also the viscous ejecta have a very small density in the polar region. Their angular mass distribution is $\propto\sin^2\theta$ \citep{Perego2017}, due to the centrifugal force in the frame that co-rotates with the disc. The wind ejecta is the only outflow emitted in the polar direction but, as said before, in BHNS mergers its mass is a very small fraction of the disc mass. Therefore we can assume that the jet looses only a negligible fraction of its energy in overcoming the ejecta and its structure is not affected. The jet kinetic energy is 
\begin{equation}
E_\mathrm{K,jet}=L_{\rm BZ}\times t_{\rm acc}
\end{equation} 
where $t_{\rm acc}$ is the duration of disc accretion
\begin{equation}
t_\mathrm{acc}=(1-\xi_\mathrm{w}-\xi_\mathrm{s})M_\mathrm{disc}/\dot{M}.
\end{equation}
The terms in parentheses represent the disc mass lost in wind and secular ejecta, thus not falling onto the BH. Then
\begin{equation}
    E_\mathrm{K,jet} = \epsilon (1-\xi_\mathrm{w}-\xi_\mathrm{s})M_\mathrm{disc}c^2\,\Omega_\mathrm{H}^2 f(\Omega_\mathrm{H}).
    \label{eq:jet_kinetic_energy}
\end{equation}
$\epsilon$ is a dimensionless constant that depends on the ratio of magnetic energy density to disc pressure at saturation \citep{Hawley2015}, on the large-scale geometry of the magnetic field and on the aspect ratio of the disc \citep{Tchekhovskoy2010}. However \citep{Shapiro2017} suggest that its value does not change significantly for different BHNS merger configurations. To fix it to a definite value, we compare the upper extremum of the SGRB energy distribution with the maximum energy that can be attained by this process. Of course the disc mass has as upper limit the total NS baryon mass, $M_\mathrm{disc}\lesssim 2 \,\mathrm{M_\odot}$, while the spin-dependent factor $\Omega_\mathrm{H}^2 f(\Omega_\mathrm{H})$ is always $<0.2$. Until today, the most energetic observed SGRB (GRB 090510) had $E_\mathrm{\gamma,iso}\sim 7.4\times 10^{52}\mathrm{erg}$ \citep{Davanzo2014}: if we assume a 10\% efficiency in converting the kinetic energy to gamma-rays and consider a jet half-aperture of $5\,\mathrm{deg}$ \citep[the typical value measured for SGRB half-opening angles, see][]{Fong2015}, we find a jet kinetic energy $E_\mathrm{K,jet}\sim 3\times 10^{51}\mathrm{erg}$. Following these arguments we set $\epsilon=0.015$, corresponding to a maximum possible jet kinetic energy of $E_\mathrm{K,jet,max} \approx 10^{52}\mathrm{erg}$.

\subsection{Jet structure}
Regardless the jet launching mechanism, it is natural to expect an angular distribution of kinetic energy per solid angle and Lorentz factor $\Gamma$ in the jet. As an educated guess, we considered the following angular distributions:
\begin{equation}\begin{split}
&\frac{dE}{d\Omega}(\theta)=E_\mathrm{c}e^{-(\theta/\theta_\mathrm{c,E})^2};
\\&\Gamma(\theta)=(\Gamma_\mathrm{c}-1)e^{-(\theta/\theta_\mathrm{c,\Gamma})^2}+1;
\end{split}\end{equation}
where we chose $E_\mathrm{c}= E_\mathrm{K,jet}/\pi\theta_\mathrm{c,E}^2$, $\theta_\mathrm{c,E}=0.1$ rad, $\Gamma_\mathrm{c}=100$, and $\theta_\mathrm{c,\Gamma}=0.2$ rad. 

In the future, this structure could be hopefully compared with real observations. Due to the negligible interaction with ejecta material, jets from BHNS mergers should bring some information on the originating region, such as the magnetic field configuration, encoded in their structure. We can argue that, if the launching conditions were similar for all the systems, the jets could present a quasi-universal structure, with only slightly different properties \citep[see][for some evidence in favour of a quasi-universal structure for jets launched by NSNS mergers]{Salafia2019}.

\subsection{GRB prompt emission}
Internal shocks and/or magnetic reconnection dissipate a typical fraction $\eta=10\%$ of the jet kinetic energy, that is radiated away giving the GRB ``prompt emission''. From \cite{Salafia2015}, the prompt isotropic equivalent energy that an observer measures at a viewing angle $\theta_\mathrm{v}$ is
\begin{equation}
    E_\mathrm{iso}(\theta_\mathrm{v})=\eta \int \frac{\delta^3}{\Gamma}\frac{dE}{d\Omega}d\Omega
,\end{equation}
where $\delta$ is the relativistic Doppler factor.
In Figure 3 from \cite{Barbieri2019} we showed the isotropic equivalent radiated energy (considering on-axis observer, $\theta_\mathrm{v}=0$ rad) dependence on the intrinsic BH properties. Obviously the jet is launched only in systems where an accretion disc forms, thus for parameter combinations resulting in a NS direct plunge no values are shown (white region). It is interesting to note that our model predicts an energy range that reproduces the observed energy range of SGRBs \citep{Davanzo2014}. 

\subsection{GRB afterglow}
After the prompt emission, the jet continues its expansion into the interstellar medium (ISM). When the jet has swept an amount of material whose rest mass energy times the jet Lorentz factor squared is comparable to the jet kinetic energy, the jet begins to decelerate. This causes the formation of a strong forward shock. Close to the shock, ISM electrons are accelerated and produce synchrotron radiation: this represents the GRB afterglow emission. Behind the forward shock (``upstream'' region) the electrons are accelerated through the Fermi process producing a non-thermal energy distribution, which is usually represented by a power-law of index $p$. We assume $p=2.3$, basing on Fermi acceleration simulations in magnetised relativistic shocks \cite{Sironi2013}. Using shock jump conditions from \cite{Blandford1976} the total energy density behind the shock can be obtained. We considered that a fraction $\epsilon_\mathrm{e}$ of this energy is given to the electrons. Similarly, a fraction $\epsilon_\mathrm{B}$ of the total energy density is given to the magnetic field (amplified by small-scale instabilities). From \cite{Beniamini2017,Nava2014}, the analysis of radio-to-GeV emission energy ratio in LGRBs suggests a typical value for $\epsilon_{\rm e} \sim 0.1$ (however we note that it is difficult to constrain this parameter from afterglow observations, due to several degeneracies). Constaints on $\epsilon_{\rm B}$ lie in the range [$10^{-4}$ - $10^{-1}$] \citep{Beniamini2016,Zhang2015,Santana2014,Granot2014}. Following \cite{Panaitescu2000} we also computed synchrotron self-absorption. We calculated the forward shock dynamics and its synchrotron emission using a modified version of the model used in \cite{D'Avanzo2018} and \cite{Ghirlanda2019}, described in \cite{Salafia2019}. We considered a constant environment density $n=10^{-3}$ cm$^{-3}$, that is typical for SGRBs with a modelled afterglow emission \citep{Fong2015} and consistent with the $n$ value estimated from the analysis of the GRB associated to GW170817 \citep{Ghirlanda2019}. A low density is in agreement with the expectation that binary merging sites are displaced from the original star-forming regions due to supernova kicks \citep[see e.g.][]{DivisionCO}.

\subsubsection{Dynamics}
In the following description we consider a jet whose axis is aligned with the $z$ axis of a spherical coordinate system. We call $\theta$ and $\phi$ the latitudinal and azimuthal angles,  respectively. The viewing angle $\theta_\mathrm{v}$ is the angle between the observer's line of sight and the jet axis. We define ``annulus'' each jet sub-region with latitudinal angle in the range [$\theta$, $\theta+d\theta$]. We do not take into account energy exchanges among adjacent annuli, that would result in lateral expansion. Therefore we consider the annuli to be independent from each other. The initial Lorentz factor and kinetic energy per unit solid angle of each annulus are, respectively, $dE/d\Omega$ and $\Gamma(0,\theta)$. The ISM mass swept from the jet per unit solid angle as a function of the radius (distance from the jet's launch site) is
\begin{equation}
\mu(R) = \frac{R^3}{3}nm_\mathrm{p},
\end{equation}
where $m_\mathrm{p}$ is the proton mass (here we are assuming the ISM to be composed only of hydrogen). Imposing energy conservation, following \cite{Granot2003} and \cite{Panaitescu2000}, we computed the shock dynamics, finding that the Lorentz factor of material right behind the shock evolves as
\begin{equation}
\Gamma(R,\theta)=\frac{\mu_0}{2\mu}\left[\sqrt{1+\frac{4\mu(dE/d\Omega~c^{-2}+\mu+\mu_0)}{\mu_0^2}}-1\right]
\end{equation}
where we defined
\begin{equation}
\mu_0(\theta)=\frac{dE/d\Omega(\theta)}{\Gamma(0,\theta)c^2}.
\end{equation}
\cite{Blandford1976} showed that the shocked material lies behind the shock in a thin layer. Assuming a uniform density distribution in the radial direction inside this layer, we can calculate its thickness $\Delta R$ imposing electron number conservation. Using shock jump conditions from \cite{Blandford1976} the number density of electrons in the shocked region is then given by
\begin{equation}
n_\mathrm{s}=n~\frac{\gamma_\mathrm{a}\Gamma+1}{\gamma_\mathrm{a}-1},
\end{equation}
with $\gamma_\mathrm{a}$ representing the post-shock adiabatic index, for which we use the formula from \cite{Peer2012}. The thickness $\Delta R$ is then
\begin{equation}
\Delta R = \frac{R(\gamma_\mathrm{a}-1)}{3(\gamma_\mathrm{a}\Gamma+1)\Gamma}.
\end{equation}
The forward shock is faster than the shocked material, and the tickness $\Delta R$ increases. The relation between the forward shock Lorentz factor $\Gamma_\mathrm{s}$ and $\Gamma$ \citep{Blandford1976} is
\begin{equation}
\Gamma_\mathrm{s}=[\gamma_\mathrm{a}(\Gamma-1)+1]\sqrt{\frac{\Gamma+1}{\gamma_\mathrm{a}(2-\gamma_\mathrm{a})(\Gamma-1)+2}}
\end{equation}

\subsubsection{Equal-arrival time surfaces}
Photons emitted in the shocked region at a given time will be detected by the observer at different times. Considering that the emitting region thickness is small with respect to the radius ($\Delta R<<R$), we can assume that photons are emitted from the shock surface. Therefore the photon arrival time is given by
\begin{equation}\label{eq:t_obs}
t_\mathrm{obs}(R,\theta_\mathrm{v},\theta,\phi)=(1+z)\int_0^{R}\frac{1-\beta_\mathrm{s}\cos{\alpha}}{\beta_\mathrm{s}c}dR,
\end{equation}
where $\beta_\mathrm{s}=\sqrt{1-\Gamma_\mathrm{s}^{-2}}$ is the shock velocity in units $c$, $z$ is the redshift and $\alpha$ is the angle between the observer line of sight and the unit vector perpendicular to the surface element, namely $\cos{\alpha}=\cos{\theta}\cos{\theta_\mathrm{v}}+\sin{\theta}\sin{\phi}\sin{\theta_\mathrm{v}}$. The shock surface brightness is given by
\begin{equation}\label{eq:brightness}
I_\nu(\nu,\theta,\phi,R)=\Delta R'j'_{\nu'}(\nu/\delta)\delta^3,
\end{equation}
where the primed quantities are calculated in the comoving frame. $j'_{\nu'}$ is the synchrotron emissivity, described in the following subsection, and $\Delta R'=\Gamma(R,\theta)\Delta R$. The term $\delta(R,\theta_\mathrm{v},\theta,\phi)=1/\{\Gamma(R,\theta)[1-\beta(R,\theta)\cos{\alpha}]\}$ is the Doppler factor of the shocked material.

We divide the jet emitting surface in sub-regions, considering $N$ bins for the latitudinal angle $\theta$ in the range [$10^{-4}$, $\pi/2$] and $M$ bins for the azimuthal angle $\phi$ in the range [$-\pi/2$, $\pi/2$]. Note that we compute the emission only from one half of the surface, as for symmetry the contribution from the other half is the same. For each sub-region we calculate $I_\nu$ at radii that correspond to a given $t_\mathrm{obs}$. We find these radii $R(\theta,\phi,t_\mathrm{obs})$ by inverting Eq.~\ref{eq:t_obs}. The total flux density at time $t_\mathrm{obs}$ is obtained by integrating over $\theta$ and $\phi$:
\begin{equation}
F_\nu(\nu,t_\mathrm{obs})=2~\frac{1+z}{d^2_\mathrm{L}}\int_0^1{d\cos{\theta}}\int_{-\pi/2}^{\pi/2}d\phi R^2I_\nu(\nu(1+z),R),
\end{equation}
where the factor $2$ takes into account the emission from the other half of the surface, using the symmetry argument previously explained. Until now we considered the emission from a single jet only. However, two jets are usually expected to be launched by the BH-disc system, in the two opposite directions along the polar axis. The receding jet is usually called ``counter-jet''. We compute the flux from the counter-jet using the same method, simply adding $\pi$ to the viewing angle.

\subsubsection{Radiation}
We adopt a shocked material synchrotron emission model similar to \cite{Panaitescu2000} and \cite{Sari1998}. We assume that the forward shock accelerates the ISM electrons generating a power law $\gamma$ distribution
\begin{equation}
\frac{dn_\mathrm{s}}{d\gamma}\propto\gamma^{-p},
\end{equation}
where $p>2$. As mentioned above, we assume that a fraction $\epsilon_\mathrm{e}$ of the total energy density $e$ behind the shock is given to the electrons:
\begin{equation}
e_\mathrm{e}=\epsilon_\mathrm{e}e=\epsilon_\mathrm{e}(\Gamma-1)n_\mathrm{s}m_\mathrm{p}c^2.
\end{equation}
Therefore we can define the minimum electron Lorentz factor \citep[``injection'',][]{Sari1998}
\begin{equation}
\gamma_m=\mathrm{max}\left[1,\frac{p-2}{p-1}(\Gamma-1)\frac{m_\mathrm{p}}{m_\mathrm{e}}\right],
\end{equation}
where $m_\mathrm{e}$ is the electron mass. Being the $\gamma$ distribution a decreasing power law, the majority of electrons have $\gamma=\gamma_\mathrm{m}$. 
Small-scale instabilities amplify the magnetic field before the shock. Again, the magnetic energy density $e_\mathrm{B}$ can be expressed as a fraction $\epsilon_\mathrm{B}$ of the total energy density $e$:
\begin{equation}
e_\mathrm{B}=\frac{B^2}{8\pi}=\epsilon_\mathrm{B}e.
\end{equation}
Following \cite{vanEerten2012} \citep[and][to take into account the ``deep newtonian'' regime]{Sironi2013} we compute the peak synchrotron emissivity of electrons ``upstream'', in the comoving frame:
\begin{equation}
j'_{\nu',\mathrm{max}}\approx0.66\frac{q_\mathrm{e}^3}{m_\mathrm{e}^2c^4}\frac{p-2}{3p-1}\frac{B\epsilon_\mathrm{e}e}{\gamma_\mathrm{m}}
\end{equation}
with $q_\mathrm{e}$ the electron charge. From Eq.~\ref{eq:brightness} the surface brightness is
\begin{equation}
I_\nu(\nu)=\delta^3\Delta R'j'_{\nu',\mathrm{max}}S(\nu'),
\end{equation}
where $\nu'=\nu/\delta$ and $S(\nu')$ is the normalized spectral shape. We compute $S(\nu')$ as a sequence of power laws, including all spectral orderings \cite[see e.g.][]{Granot2002}. The different power laws are connected at break frequencies: $\nu_\mathrm{m}$, $\nu_\mathrm{c}$, $\nu_\mathrm{a}$ and $\nu_\mathrm{ac}$.
$\nu_\mathrm{m}$ is the synchrotron frequency related to $\gamma_\mathrm{m}$,
\begin{equation}
\nu_\mathrm{m}=\frac{\gamma^2_\mathrm{m}q_\mathrm{e}B}{2\pi m_\mathrm{e}c}.
\end{equation}
$\nu_\mathrm{c}$ is the synchrotron frequency related to $\gamma_\mathrm{c}$, the electron Lorentz factor above which they loose energy through synchrotron emission on a timescale shorter than the dynamical timescale of expansion (over which ``fresh'' electrons are injected upstream):
\begin{equation}
\gamma_\mathrm{c}=\frac{6\pi m_\mathrm{e}c^2\Gamma\beta}{\sigma_\mathrm{T}B^2R},
\end{equation}
with $\sigma_\mathrm{T}$ the Thomson cross section.

If $\nu_\mathrm{m}<\nu_\mathrm{c}$ the majority of electrons have a Lorentz factor smaller than $\gamma_\mathrm{c}$, thus their synchrotron emission energy loss happens on a timescale larger than the expansion timescale. This regime is defined ``slow cooling''. If instead $\nu_\mathrm{m}>\nu_\mathrm{c}$ the electrons loose their energy faster than the shocked region is refilled with fresh electrons. This regime is defined ``fast cooling''. Emitted photons can be re-absorbed by electrons before leaving the shocked region in free-free transitions: this phenomenon is called synchrotron self-absorption. Below the frequency $\nu_\mathrm{a}$ the emission is self-absorbed \citep{Panaitescu2000}. For fast cooling, another break frequency ($\nu_\mathrm{ac}<\nu_\mathrm{a}$) in the self-absorbed part of the spectrum arises, due to the non-homogeneous distribution of electrons at different cooling stages \citep{Granot2000}.

\begin{figure*}
    \centering
    \includegraphics[]{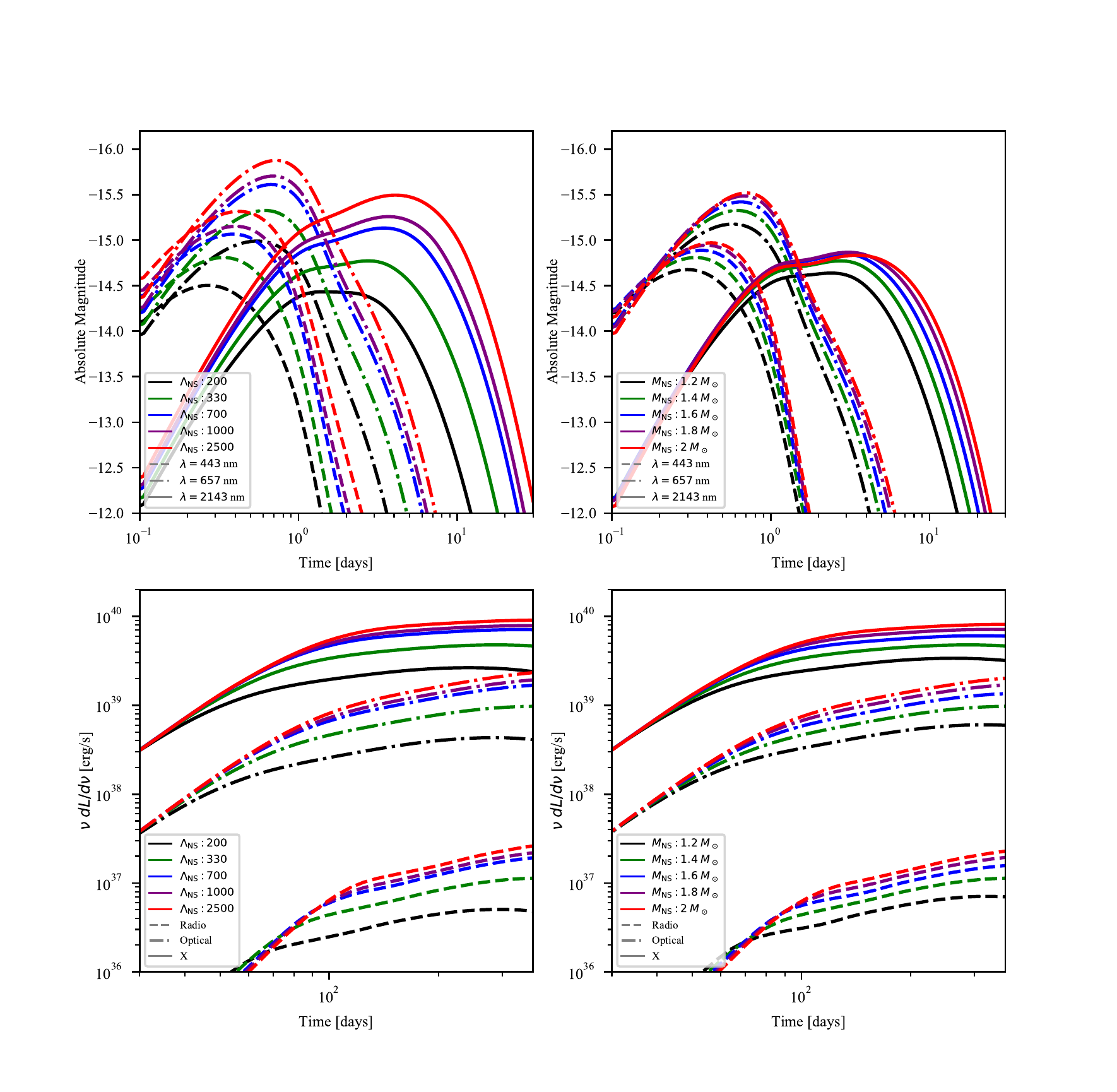}
    \caption{Kilonova (top) and GRB afterglow (bottom) light curve dependence on NS tidal deformability (left - for $M_{\rm NS}=1.4\msun$) and mass (right - for $\Lambda = 330$), considering a BH with $M_\mathrm{BH}=3~M_\odot$ and $\chi_\mathrm{BH}=0.5$. For the kilonova we show the absolute magnitude \textit{vs} time, while for the GRB afterglow we show the quantity $\nu~dL/d\nu$ \textit{vs} time. Linestyles indicate different wavelengths (for the kilonova, listed in the legend) or frequencies (for the GRB): $1.4\,\mathrm{GHz}$ (``Radio''), $4.6\times 10^{14}\,\mathrm{Hz}$ (``Optical'') and $2.4\times 10^{17}\,\mathrm{Hz}=1\,\mathrm{keV}/h$ (``X-ray''). Note that kilonova light curves are plotted from $0.1$ to $30$ days, while GRB afterglow light curves are plotted from $30$ days to $1$ year. 
    }
    \label{fig:lc_3_5}
\end{figure*}

\section{Kilonova and GRB afterglow light curve dependence on NS properties}\label{sec:kn+grb}

\begin{figure*}
    \centering
    \includegraphics[]{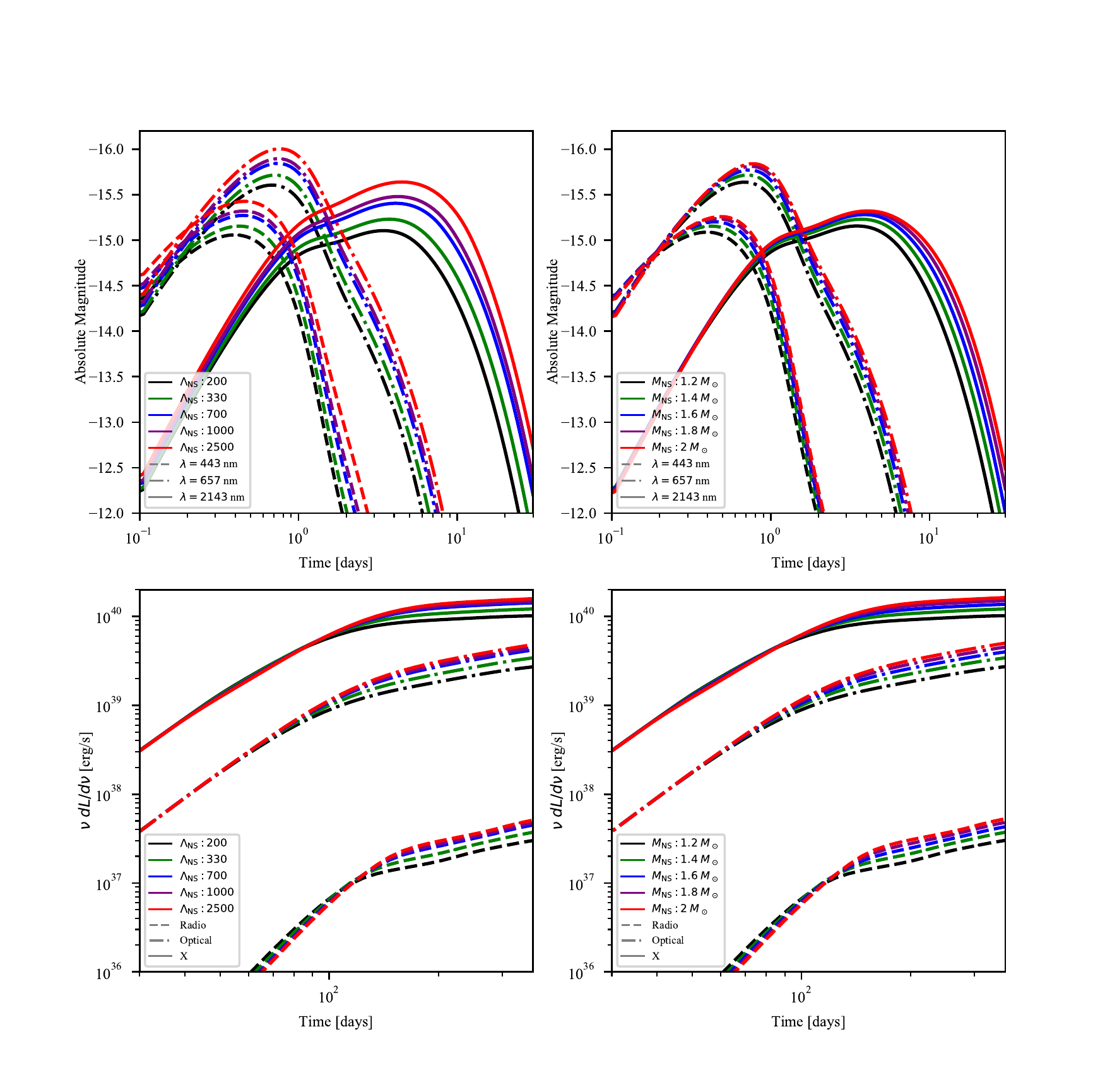}
    \caption{Same as Fig. \ref{fig:lc_3_5}, considering a BH with $M_\mathrm{BH}=3~M_\odot$ and $\chi_\mathrm{BH}=0.8$.}
    \label{fig:lc_3_8}
\end{figure*}

In this section we analyse the fundamental dependencies of kilonova and GRB afterglow light curves on the NS properties ($M_\mathrm{NS}$ and $\Lambda_\mathrm{NS}$). In particular, for each set of BH properties, we first fixed the NS mass to $1.4\msun$  and explored a set of values for $\Lambda_\mathrm{NS}$, then we did the opposite, fixing $\Lambda_\mathrm{NS}$ to $330$ and varying $M_\mathrm{NS}$. We stress that in Figs.~\ref{fig:lc_3_5}-\ref{fig:lc_6_8} we considered NS parameter pairs independently from existing EoS. Instead in Figs.~\ref{fig:eos_3_5}-\ref{fig:eos_6_8} we show light curves obtained using NS parameter pairs consistent with two EoS.

\subsection{Kilonova}

\begin{figure*}
    \centering
    \includegraphics[]{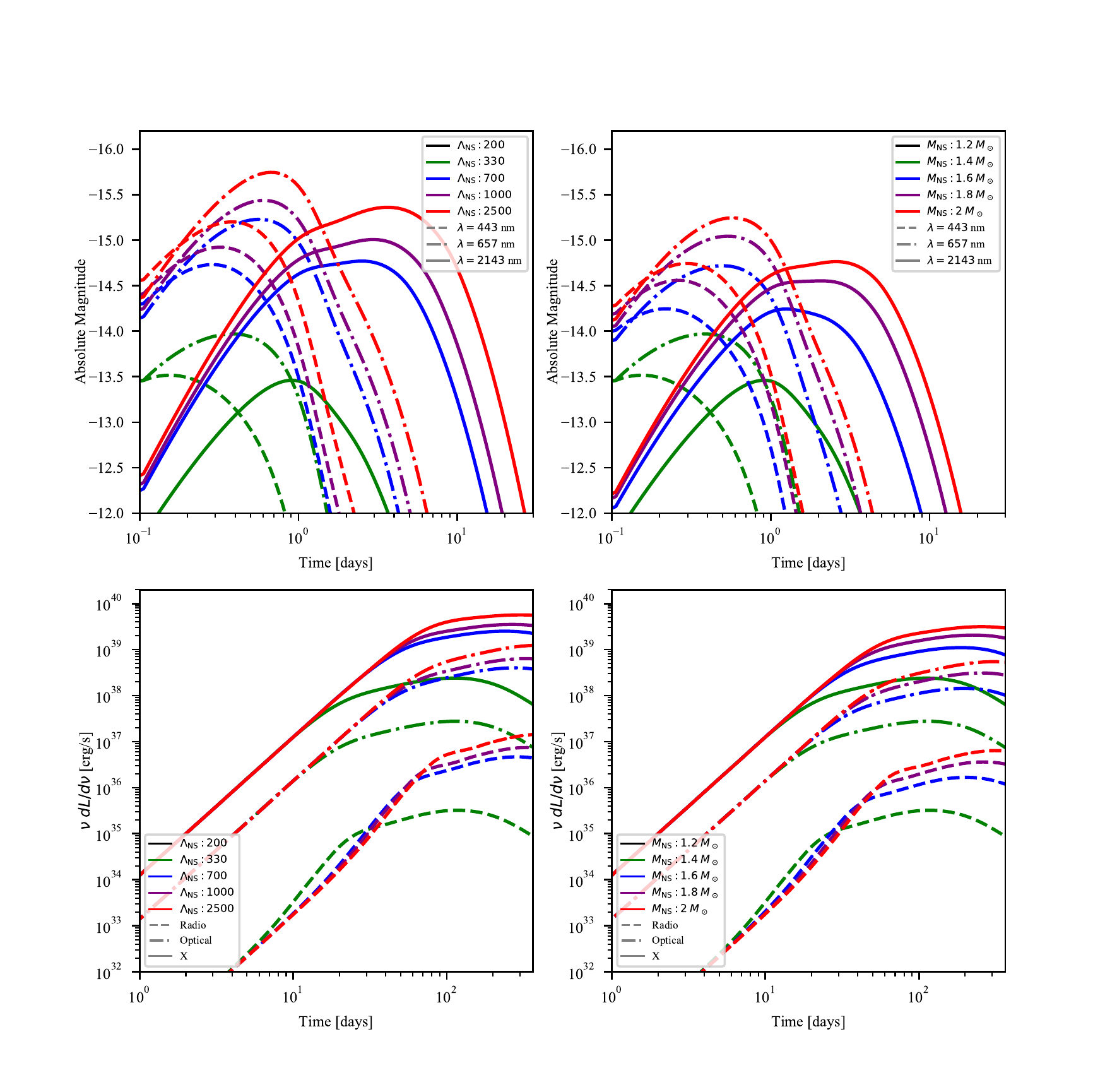}
    \caption{Same as Fig. \ref{fig:lc_3_5}, considering a BH with $M_\mathrm{BH}=6~M_\odot$ and $\chi_\mathrm{BH}=0.5$. Note that in this case the GRB afterglow light curves are plotted from $1$ day to $1$ year.}
    \label{fig:lc_6_5}
\end{figure*}

\begin{figure*}
    \centering
    \includegraphics[]{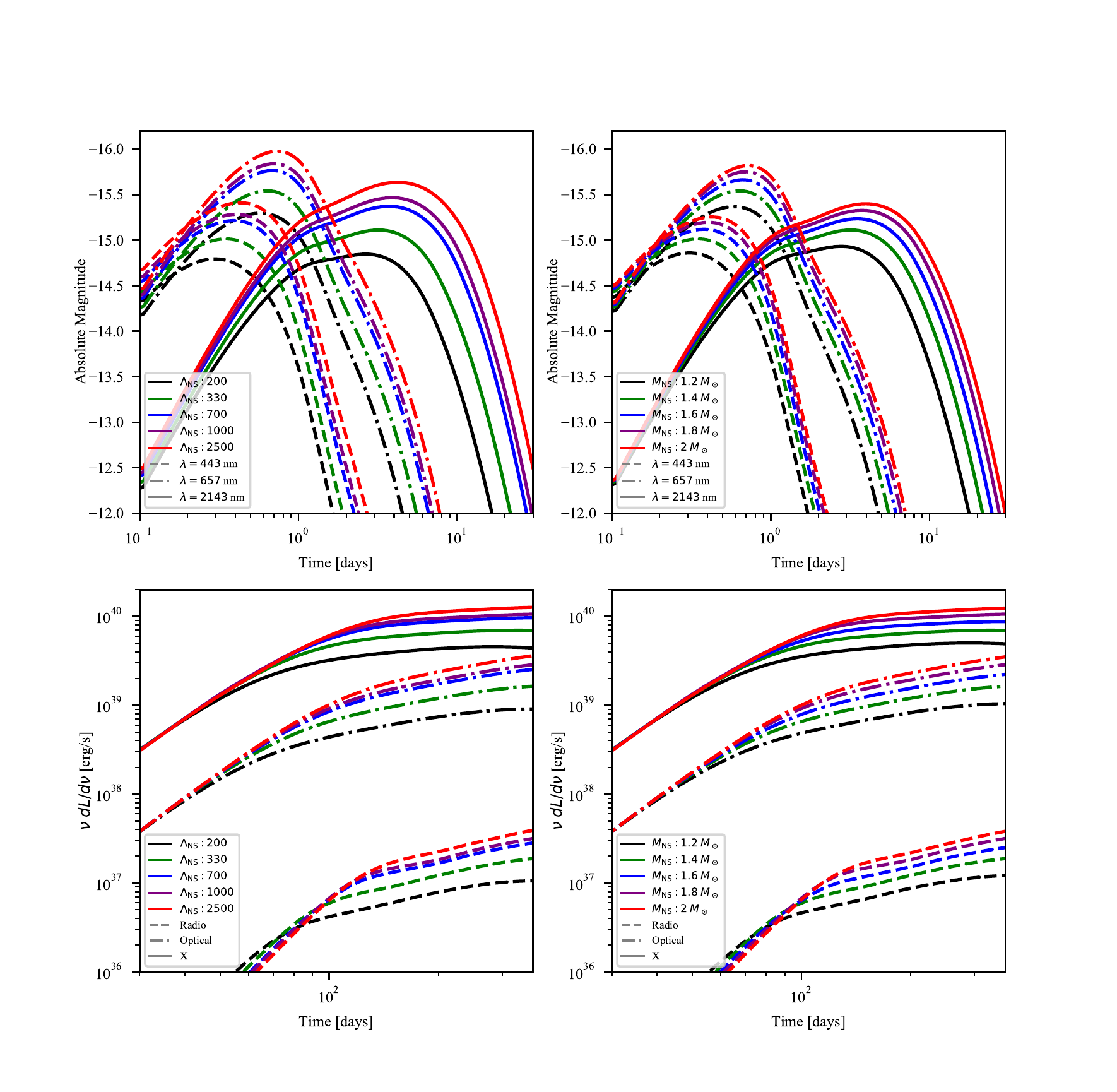}
    \caption{Same as Fig. \ref{fig:lc_3_5}, considering a BH with $M_\mathrm{BH}=6~M_\odot$ and $\chi_\mathrm{BH}=0.8$.}
    \label{fig:lc_6_8}
\end{figure*}

In the top panels of Figs.~\ref{fig:lc_3_5}-\ref{fig:lc_6_8} we show the kilonova light curves as function of $\Lambda_\mathrm{NS}$ (left-hand panel) and $M_\mathrm{NS}$ (right-hand panel). For a given NS mass, the larger $\Lambda_\mathrm{NS}$ the brighter the kilonova. Indeed, as explained above, fixing $M_\mathrm{NS}$ and increasing the NS tidal deformability leads to the production of more massive ejecta, giving a more luminous kilonova (Eq.~\ref{eq:kn_lum}). For a given $\Lambda_\mathrm{NS}$, the more massive the NS the brighter the kilonova. Indeed, increasing the NS mass related to a certain $\Lambda_\mathrm{NS}$ amounts to moving towards stiffer EoS (as can be easily understood from Fig.~\ref{fig:lambda-m}). As explained above, a stiffer EoS leads to the production of more massive ejecta, again giving a more luminous kilonova.

This is the general trend, but there is an exception. In the top-right panel of Fig.~\ref{fig:lc_3_5} (light curves with fixed $\Lambda_\mathrm{NS}=330$) we find a different behaviour. The KN light curves corresponding to $M_\mathrm{NS}=1.6~M_\odot$ (blue) and $M_\mathrm{NS}=1.8~M_\odot$ (purple) are not always dimmer than that corresponding to $M_\mathrm{NS}=2~M_\odot$ (red). The former light curve is brighter until $\approx~$hours for $B$ and $r$ bands and $\approx~4$ days for $K$ band. From Fig.~\ref{fig:ejecta_m3_s5} we see that for $\Lambda_\mathrm{NS}=330$ the dynamical ejecta mass decreases for $M_\mathrm{NS}\geq1.8~\msun$ instead of increasing. The disc mass instead always increases\footnote{This may be due to the different dependence of the disc and dynamical ejecta mass on the NS compactness (Eqs.~\ref{eq:mass-out} and \ref{eq:Mdyn}), but we caution again that the two fitting formulae are based on different datasets, and that the one from \cite{Kawaguchi2016} is not calibrated in this part of the parameter space.} with more massive NSs at fixed $\Lambda_\mathrm{NS}$. Therefore at early times, where the major contribution is from dynamical ejecta, the $M_\mathrm{NS}=2~M_\odot$ light curve is dimmer than the $M_\mathrm{NS}=1.6-1.8~M_\odot$ ones, because it corresponds to less massive dynamical ejecta. Instead at late times, when wind and secular ejecta (originating from the disc) emission is dominant, the light curve is brighter, because it corresponds to a more massive disc.

\subsection{GRB afterglow}
The same arguments hold for GRB afterglow light curves. In the bottom panels of Figs.~\ref{fig:lc_3_5}--\ref{fig:lc_6_8} we show the time evolution of $\nu\,dL/d\nu$ for three representative frequencies: $1.4\,\mathrm{GHz}$ (``Radio''), $4.6\times 10^{14}\,\mathrm{Hz}$ (``Optical'') and $2.4\times 10^{17}\,\mathrm{Hz}=1\,\mathrm{keV}/h$ (``X-ray''). As evident from Fig.~6 in \cite{Barbieri2019}, the GRB afterglow light curve strongly depends on the viewing angle. Here we fixed $\theta_\mathrm{view}=30^\circ$, indicated in \cite{Schutz2011} as the most probable viewing angle for GW signal detections from binary mergers. In bottom left panels we see that, for a given $M_\mathrm{NS}$, the larger the NS tidal deformability the brightest the GRB afterglow. In bottom right panels we see that increasing $M_\mathrm{NS}$ for a given $\Lambda_\mathrm{NS}$ leads to brighter GRB afterglows.

We computed both kilonova and GRB afterglow light curves also considering a BH spin $\chi_\mathrm{BH}=0.3$, but we do not show them for briefness. We just discuss here the effect of a smaller $\chi_\mathrm{BH}$ on the light curves. As explained before, BHNS with low-spinning BHs give rise to an EM counterpart if the mass ratio is small and/or the NS tidal deformability is large. The corresponding ejecta mass is smaller with respect to the case with larger spin, therefore we expect the light curves to be dimmer. Indeed we find that kilonovae for $\chi_\mathrm{BH}=0.3$ are generally $\sim~0.5$ -- 1 mag below the case with $\chi_\mathrm{BH}=0.5$, and for configurations with small NS mass and/or small deformability they are not produced at all. The same holds for GRB afterglows, where for $\chi_\mathrm{BH}=0.3$ the quantity $\nu\,dL/d\nu$ is generally $\sim~5$ -- 20 times smaller than the $\chi_\mathrm{BH}=0.3$ case.

\begin{figure*}
    \centering
    \includegraphics[]{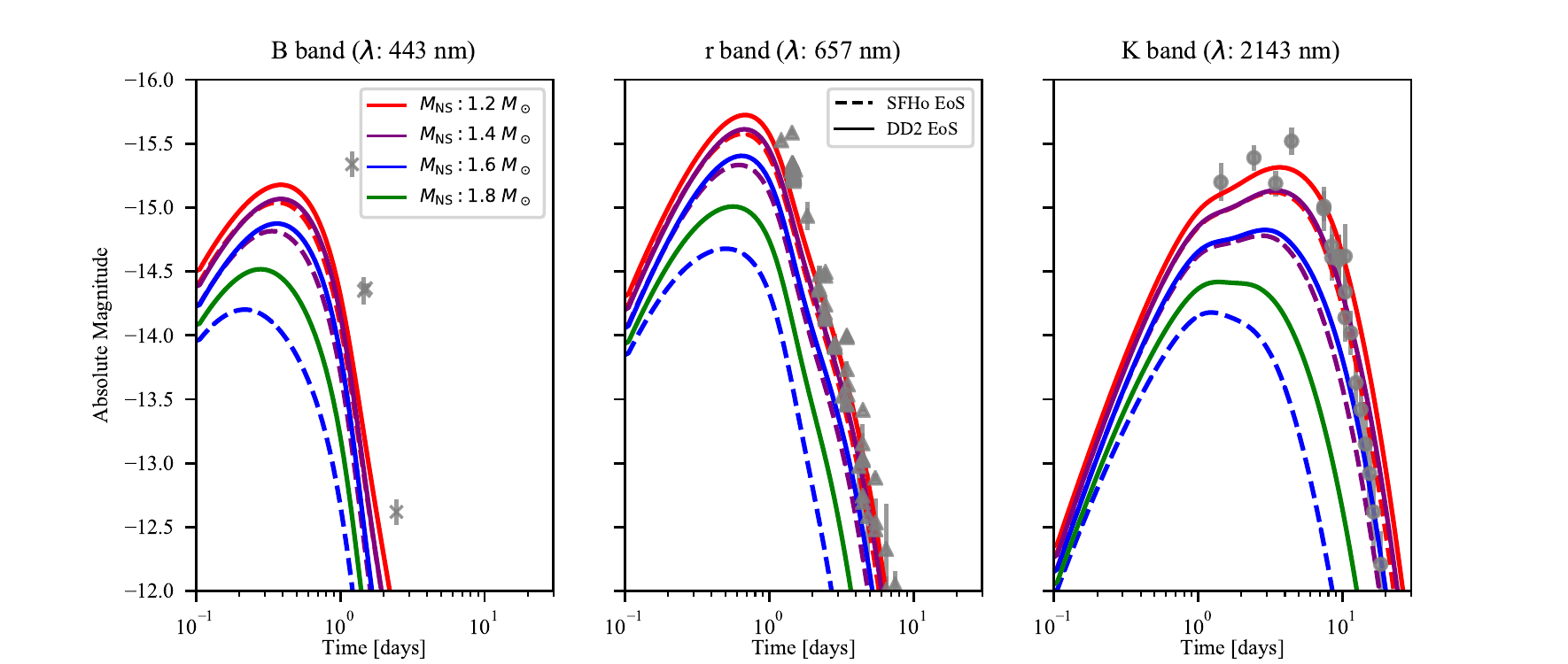}
    \caption{Kilonova light curve dependence on NS mass and EoS, considering a BH with $M_\mathrm{BH}=3~M_\odot$ and $\chi_\mathrm{BH}=0.5$. We show the absolute magnitude VS time. Linestyles indicate the two different adopted EoS reported in the central legend, while colours indicate different NS masses. We show in gray observations of the kilonova associated with GW170817.}
    \label{fig:eos_3_5}
\end{figure*}

\subsection{Selecting the EoS}\label{sec:sel_eos}
In order to compute these light curves we considered different combinations of $M_\mathrm{NS}$ and $\Lambda_\mathrm{NS}$. However, as explained above, only certain parameter pairs are consistent with an EoS. Therefore, by selecting a set of EoS consistent with the observation of GW170817, we can reduce the span in absolute magnitude for expected light curves. \cite{Radice2018_3} found joint constraints on NS EoS from GW170817 multimessenger observations. As shown in their Fig.~2, the upper and lower limits on the system tidal parameter $\tilde{\Lambda}$ are close to the DD2 \citep{DD2,DD2_2} and SFHo EoS, respectively. Green light curves in Figs.~\ref{fig:lc_3_5}--\ref{fig:lc_6_8} represent our results for the SFHo EoS (giving $\Lambda_\mathrm{NS}\approx330$ for a $1.4~M_\odot$ NS), while blue light curves represent those for the DD2 EoS (giving $\Lambda_\mathrm{NS}\approx700$ for a $1.4~M_\odot$ NS and $\Lambda_\mathrm{NS}\approx330$ for a $1.6~M_\odot$ NS). Therefore, using the above constraints on the NS EoS, for given $M_\mathrm{BH}$, $\chi_\mathrm{BH}$ and $M_\mathrm{NS}$ (or, equivalently, $\Lambda_\mathrm{NS}$) our predictions on kilonova light curves fall in a narrow absolute magnitude interval.

\begin{figure*}
    \centering
    \includegraphics[]{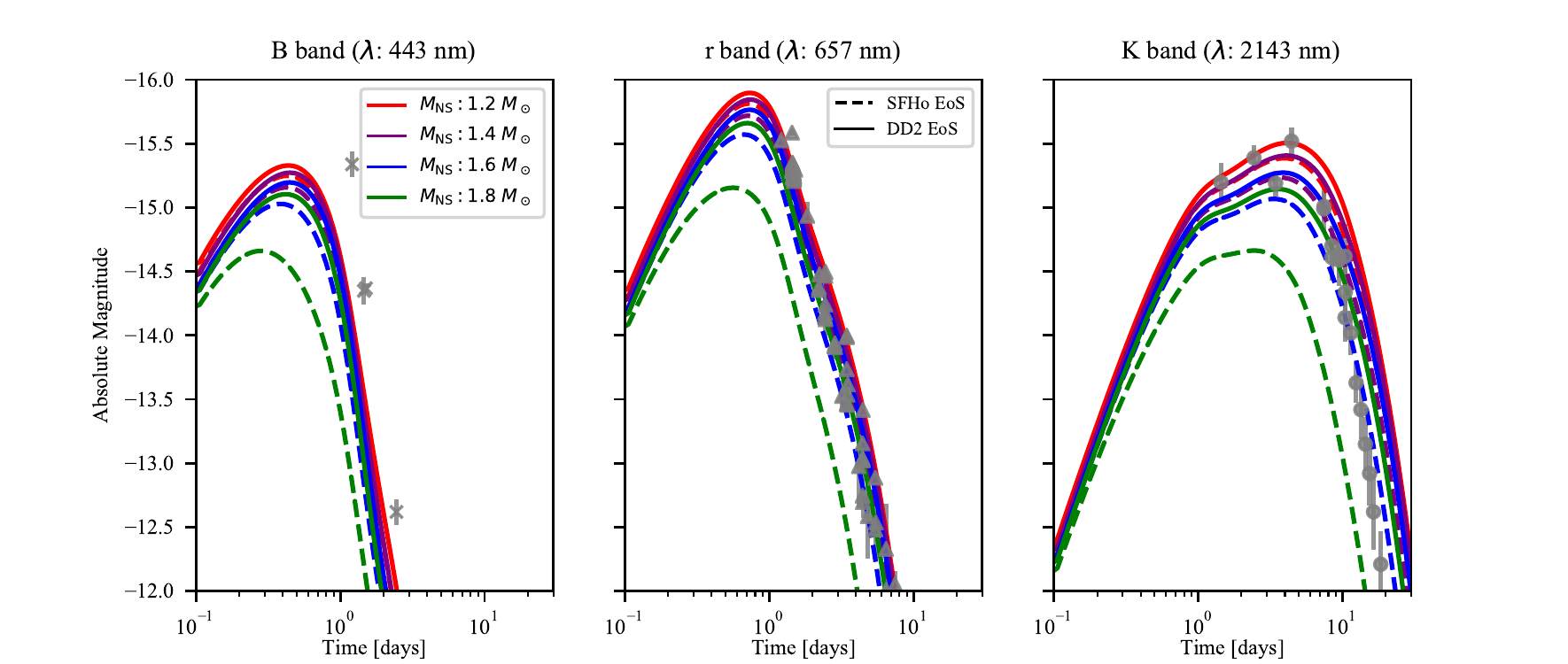}
    \caption{Same as Fig. \ref{fig:eos_3_5}, considering a BH with $M_\mathrm{BH}=3~M_\odot$ and $\chi_\mathrm{BH}=0.8$.}
    \label{fig:eos_3_8}
\end{figure*}

\begin{figure*}
    \centering
    \includegraphics[]{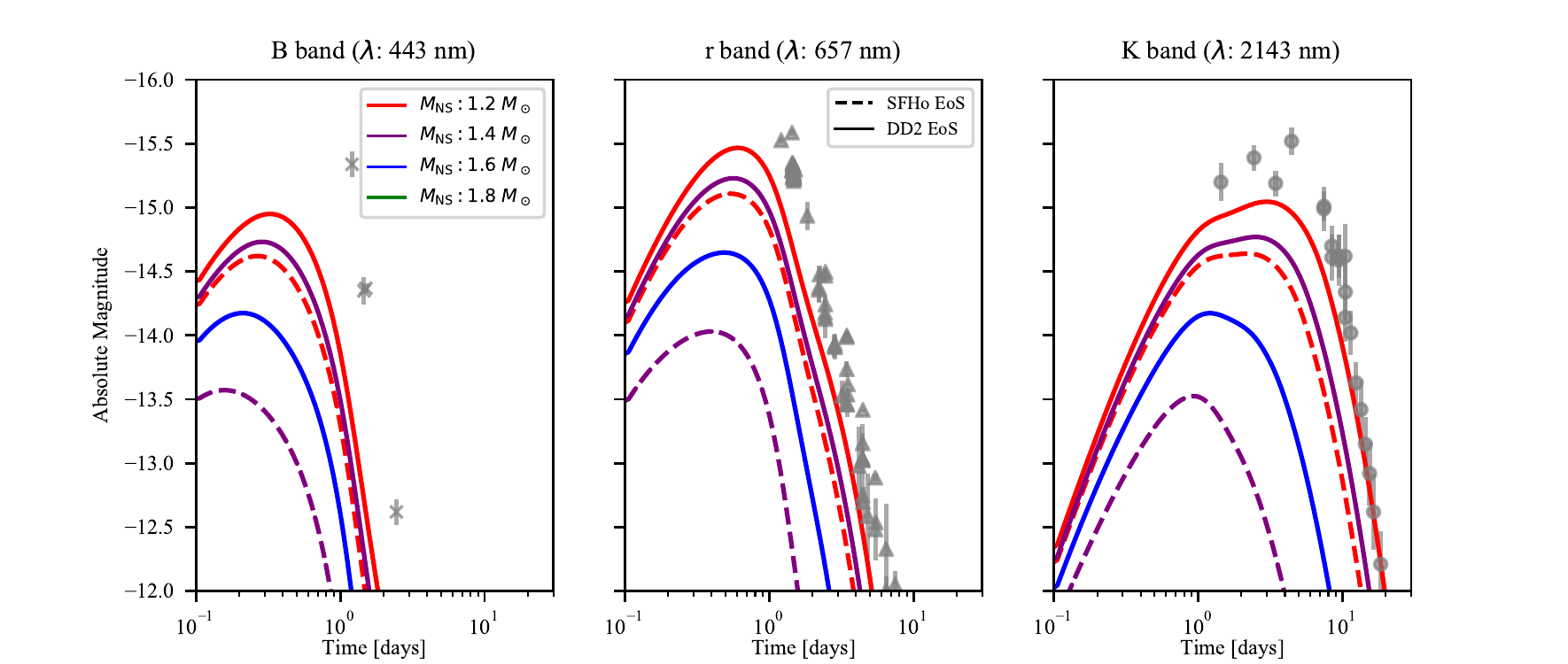}
    \caption{Same as Fig. \ref{fig:eos_3_5}, considering a BH with $M_\mathrm{BH}=6~M_\odot$ and $\chi_\mathrm{BH}=0.5$.}
    \label{fig:eos_6_5}
\end{figure*}

\begin{figure*}
    \centering
    \includegraphics[]{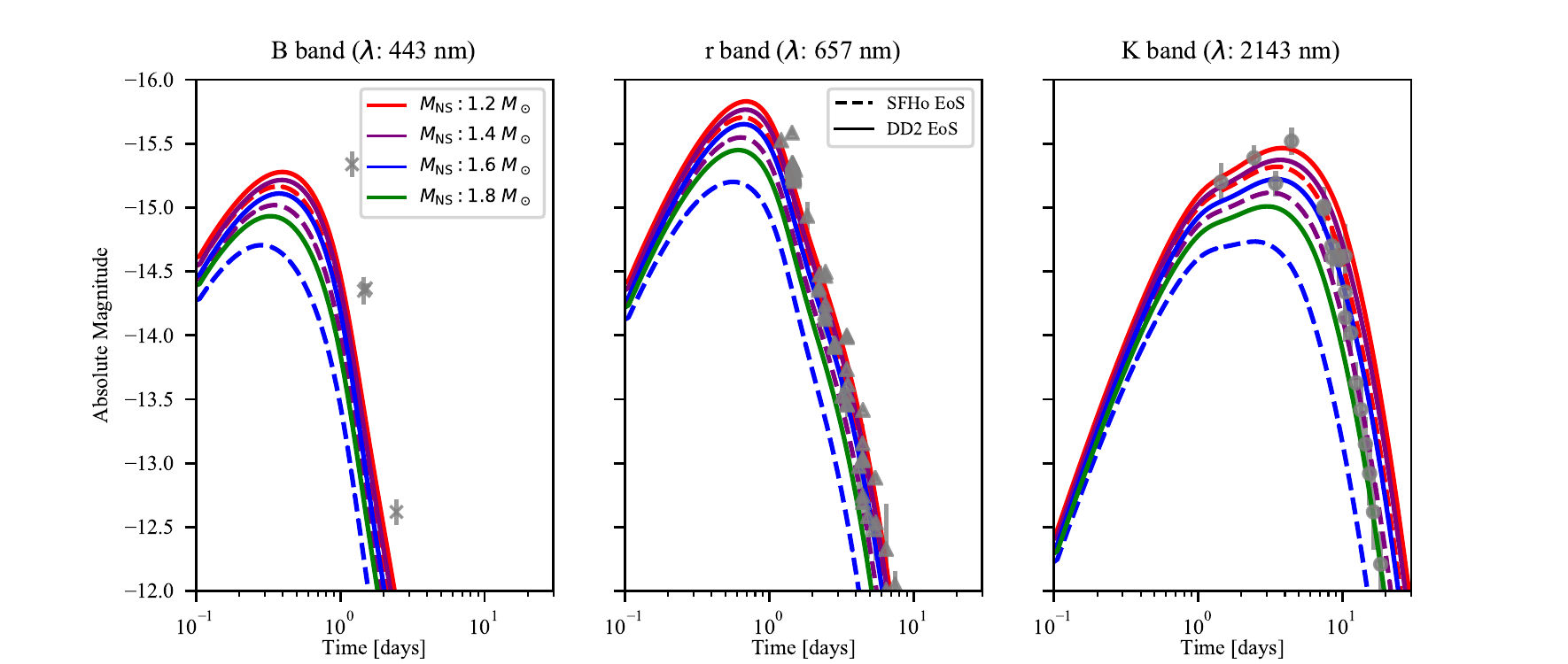}
    \caption{Same as Fig. \ref{fig:eos_3_5}, considering a BH with $M_\mathrm{BH}=6~M_\odot$ and $\chi_\mathrm{BH}=0.8$.}
    \label{fig:eos_6_8}
\end{figure*}

The previous arguments on the optimal condition to produce bright EM counterparts hold when we freely explore the $M_\mathrm{NS}$--$\Lambda_\mathrm{NS}$ parameter space. If we instead limit our analysis along a given EoS (as if nature would select a universal relation), then each NS mass is linked to a $\Lambda_\mathrm{NS}$. As shown in Fig.~\ref{fig:lambda-m}, the largest values for the NS tidal deformability correspond to low-mass NSs. Therefore, by exploring the NS parameter space moving only on EoS lines, we find that the brightest EM counterparts are produced in binaries with a low-mass NS. This is shown in Figs.~\ref{fig:eos_3_5}-\ref{fig:eos_6_8}, where for simplicity we plot only the kilonova emission. Here we see that, for a given EoS, light curves get dimmer with increasing NS mass.

\subsubsection{Comparing BHNS kilonova lightcurves with AT2017gfo-GW170817 counterpart}{}
As a comparison, in Figs.~\ref{fig:eos_3_5}-\ref{fig:eos_6_8} we show the observations of the kilonova associated with GW170817 (AT2017gfo). We took data points of apparent magnitudes from \cite{Villar2017}, we corrected for interstellar extinction following \cite{Cardelli1989} (de-reddening), then we calculated the absolute magnitudes assuming the GW170817 source luminosity distance $d_\mathrm{L}=40$ Mpc \citep{GW170817}. It is interesting to note that the general light curve behaviour and peak times are similar. However, expecially in the B band, the expected kilonova light curves for all the considered BHNS configurations are always dimmer with respect to AT2017gfo. The emission in this band is principally due to high $Y_\mathrm{e}$ ejecta. Therefore this difference can be explained with the smaller mass of the wind ejecta (the one with larger $Y_\mathrm{e}$) in BHNS merger compared to NSNS merger. Indeed, as discussed in \S~\ref{sec:klc}, only in the latter case an intermediate supra- or hypermassive NS can form before collapsing to a BH. This transient object produces an intense neutrino wind that interacts with the ejecta, increasing the electron fraction. As far as the r and K bands are concerned, there are some BHNS configurations for which the expected kilonova light curves are consistent with AT2017gfo, expecially for the cases with $\chi_\mathrm{BH}=0.8$ (Figs.~\ref{fig:eos_3_8}-\ref{fig:eos_6_8}). This is very important, showing that having only an EM observation of a kilonova without a GW signal, r and K bands measurements do not allow to distinguish between a NSNS or a BHNS merger, while B band observations can break this degeneracy.

\subsection{Limiting cases}
In our study, Fig. \ref{fig:lc_6_5}, for a BH with $M_{\rm BH}=6\msun$ and $\chi_{\rm BH}=0.5$, shows the largest spread in the light curves. First, no EM counterpart is produced for $\Lambda_\mathrm{NS}=200-M_\mathrm{NS}=1.4~M_\odot$ and $\Lambda_\mathrm{NS}=330-M_\mathrm{NS}=1.2~M_\odot$. As indicated in Fig.~\ref{fig:ejecta_m6_s5}, for these  combinations the NSs plunge directly onto the BH.  Again from Fig.~\ref{fig:ejecta_m6_s5}, we see that the wide spread
in the light curves can be explained by the large variability of dynamical ejecta and disc mass in the explored region: compared to other BH configurations, in this case we are testing the NS parameter region close to the direct plunge. The green light curve ($\Lambda_\mathrm{NS}=330-M_\mathrm{NS}=1.4~M_\odot$) is very dim and peaks early with respect to the others, due to the small amount of matter powering the emission.

\section{Energy radiated in kilonova}\label{sec:energy_kn}
In this section we calculate the total energy emitted in the  kilonova. We compute this quantity as 
\begin{equation}
E_\mathrm{KN} = \int {L_\mathrm{bol,TOT}~dt},     
\end{equation}
where
\begin{equation}
L_\mathrm{bol,TOT}=L_\mathrm{bol,dynamical}+L_\mathrm{bol,wind}+L_\mathrm{bol,secular}.
\end{equation}
We compute $L_\mathrm{bol,wind}$ and $L_\mathrm{bol,secular}$ summing Eq.~\ref{eq:kn_lum} over all the angular bins. Instead 
\begin{equation}
L_\mathrm{bol,dynamical} = L_\mathrm{bol,rad}+L_\mathrm{bol,lat}, 
\end{equation} 
where $L_\mathrm{bol,rad}$ (from Eq.~\ref{eq:lum_kn_rad}) refers to radial emission and $L_\mathrm{bol,lat}$ is the integral of Eq.~\ref{eq:lum_kn_lat} over the velocity distribution of dynamical ejecta.

\begin{figure}
    \centering
    \includegraphics[]{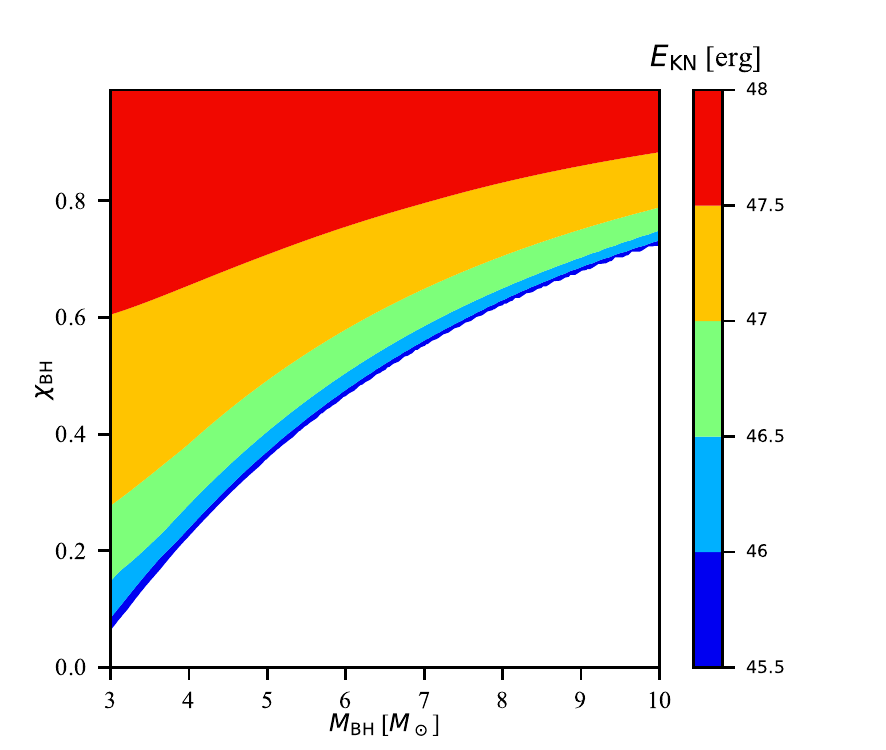}
    \caption{Total energy emitted in kilonova in the $M_\mathrm{BH}-\chi_\mathrm{BH}$ parameter space. We assumed a NS with $M_\mathrm{NS}=1.4~M_\odot$ and $\Lambda_\mathrm{NS}=330$, corresponding to SFHo EoS).}
    \label{fig:Ekn_BH}
\end{figure}

In Fig. \ref{fig:Ekn_BH}, for a given NS ($M_\mathrm{NS}=1.4\msun$ and $\Lambda_\mathrm{NS}=330$, corresponding to SFHo EoS), we show the total energy released in the kilonova in the  parameter space $M_\mathrm{BH}-\chi_\mathrm{BH}$. It is apparent that more energetic kilonovae are produced by faster spinning and/or less massive BHs. This $E_\mathrm{KN}$ dependence on the BH properties mirrors the $M_\mathrm{out}$ dependence. 

\begin{figure}
    \centering
    \includegraphics[]{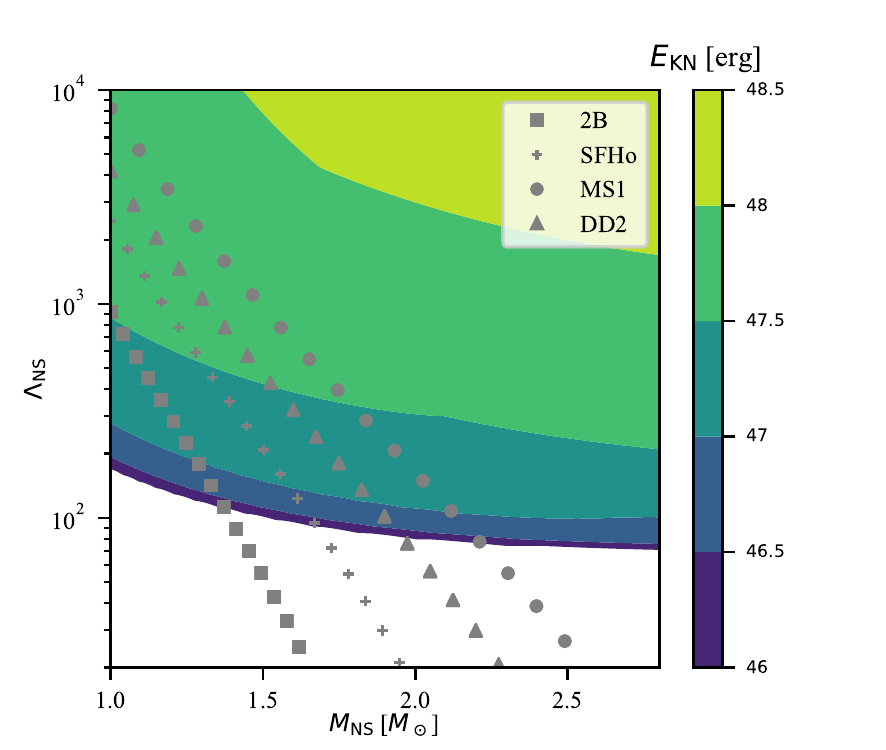}
    \caption{Total energy emitted in kilonova in the $M_\mathrm{NS}-\Lambda_\mathrm{NS}$ parameter space, for a BH with $M_\mathrm{BH}=3~M_\odot$ and $\chi_\mathrm{BH}=0.5$.}
    \label{fig:Ekn_NS_3_5}
\end{figure}

\begin{figure}
    \centering
    \includegraphics[]{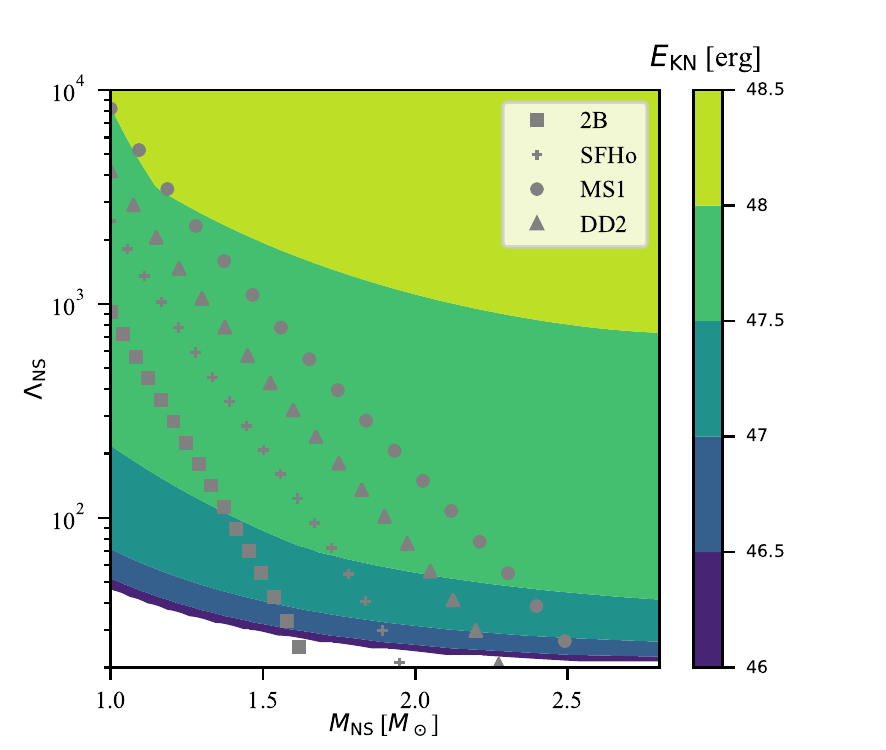}
    \caption{Same as Fig. \ref{fig:Ekn_NS_3_5}, for a BH with $M_\mathrm{BH}=3~M_\odot$ and $\chi_\mathrm{BH}=0.8$.}
    \label{fig:Ekn_NS_3_8}
\end{figure}

\begin{figure}
    \centering
    \includegraphics[]{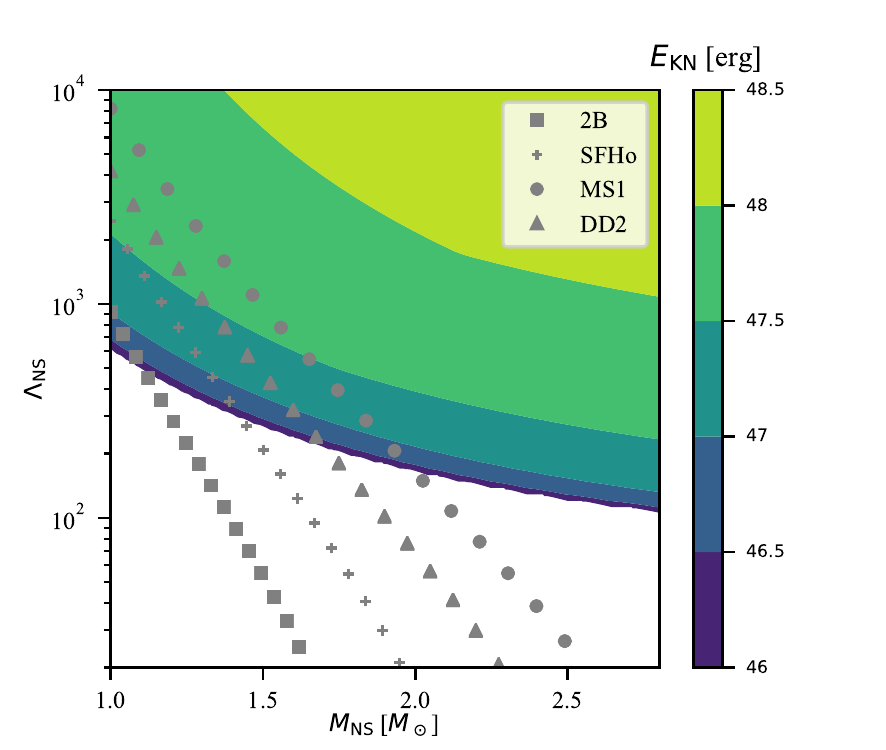}
    \caption{Same as Fig. \ref{fig:Ekn_NS_6_5}, for a BH with $M_\mathrm{BH}=6~M_\odot$ and $\chi_\mathrm{BH}=0.5$.}
    \label{fig:Ekn_NS_6_5}
\end{figure}

\begin{figure}
    \centering
    \includegraphics[]{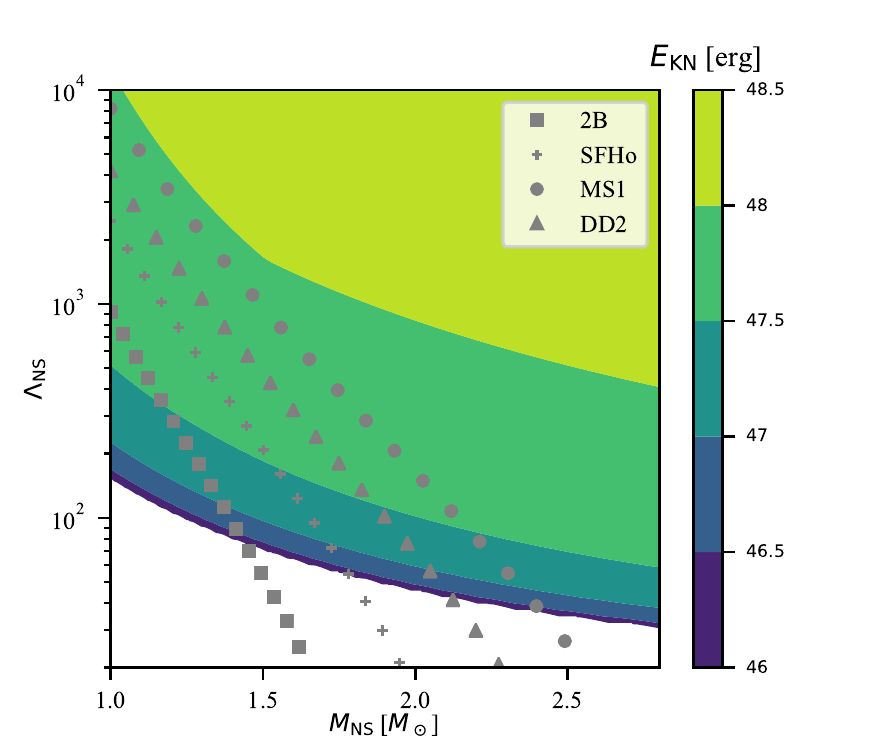}
    \caption{Same as Fig. \ref{fig:Ekn_NS_6_8}, for a BH with $M_\mathrm{BH}=6~M_\odot$ and $\chi_\mathrm{BH}=0.8$.}
    \label{fig:Ekn_NS_6_8}
\end{figure}

In Figs.~\ref{fig:Ekn_NS_3_5}--\ref{fig:Ekn_NS_6_8} we show the total energy emitted in kilonova in the $M_\mathrm{NS}$--$\Lambda_\mathrm{NS}$ plane, for different BH parameters. Again, a clear correspondence with the $M_\mathrm{out}$ dependence on NS parameters is apparent. 

It is interesting to note that the total energy radiated in 
the kilonova is always $\sim~10^{-5}$ times the ejecta rest mass energy (and $\sim10^{-4}$ times the ejecta kinetic energy, which is $\sim$ few percents of the rest mass energy). This energy transformation efficiency is consistent with the typical one for nuclear processes. We note that this efficency is much lower compared to the energy emitted in GWs.  Indeed, the total energy emitted in GWs can reach $\sim 5~\msun c^2$ ($\sim10^{55}$ erg) in BHBH mergers and $\sim0.05~\msun c^2$ ($\sim10^{53}$ erg) in NSNS merger \citep{Zappa2018,CatalogoGW}.

\section{Conclusion}
BHNS binary mergers are the next multi-messenger astronomy sources to be detected. While a firm detection of a GW signal from these sources is still awaited at the time of writing \citep[although there has been a promising candidate, see][]{GCN190426c}, one of the possible EM counterparts from this family of mergers could have already been observed, nestled among the short GRB population \citep[][and references therein]{Gompertz2018,Rossi2019}.

In this paper, we presented a composite model to compute the BHNS merger EM counterpart light curves. During the final phase of the inspiral, just before crossing the BH event horizon, the NS can be partially disrupted, leaving neutron-rich material outside the BH. Some of this material is gravitationally unbound (``dynamical'' ejecta), while the rest forms an a
ccretion disc around the remnant BH created after the merger. From this disc two additional outflows are produced: the ``wind'' ejecta and ``secular'' ejecta. Finally, accretion onto the BH can power the launch of a relativistic jet. These outflows produce the EM counterparts: the non-relativistic ejecta power the kilonova emission, while the relativistic jet produces the GRB afterglow emission. For simplicity, in this work we considered only these two emission processes, leaving out, for instance, the jet prompt emission and the kilonova radio remnant \citep[discussed in some detail in][]{Barbieri2019}. Therefore the NS tidal disruption is crucial in order to produce EM counterparts in a BHNS merger.

We studied dependence of the dynamical ejecta and disc masses on the NS properties, namely the mass $M_\mathrm{NS}$ and dimensionless tidal deformability parameter $\Lambda_\mathrm{NS}$, for fixed BH properties (Figs. \ref{fig:ejecta_m3_s5}-\ref{fig:ejecta_m6_s8}). We found that NSs never suffer a total tidal disruption. Indeed for one of the stiffest physically motivated EoS (MS1) the mass remaining outside the remnant BH is $\lesssim40$ \% of the NS mass.

In the computation of the kilonova emission from dynamical, wind and secular ejecta we take into account the expected anisotropies
emerging from the eject in the form of a crescent (see \S \ref{sec:kilonova}). We also modeled GRB afterglow emission accounting for anisotropies in the relativistic jet, for the angular distribution of both the energy and the Lorentz factor (see \S \ref{sec:grb_afterglow}).

In \S \ref{sec:kn+grb} we presented different sets of light curves. For each set we fixed the BH mass and spin, and we studied their dependence on the NS mass and tidal deformability, by varying one at a time the two quantities. 
EM lightcurves from BHNS mergers display a large degeneracy introduced by different combinations of binary parameters. Therefore we stress that, through the EM multi-wavelength observations alone, it is not possible to constrain the intrinsic binary parameters. By contrast,  this degeneracy can be broken by performing a multi-messenger analysis with joint GW and EM signals to infer both BH and NS properties, as suggested in \cite{Barbieri2019}. 

Using the constraints on the NS EoS from GW170817 \citep{Radice2018_3} our modeling of the kilonova light curves shows that the emission falls in a narrow range of absolute magnitudes, and that the brightest EM counterparts are associated to BHNS binaries with a low-mass NS of 1-1.2 $\msun$. The light curves behaviour and peak times are not dissimilar to NSNS mergers, except in the B band. The lack of an intense neutrino wind, originating from the supra/hyper-massive NS that could form in NSNS coalescences, leads to a dimming of the blue component of the kilonova emission in BHNS mergers, should the kilonova from NSNS binaries display a universal behaviour. Our work provides light curves that may serve as guide to identify EM counterparts of BHNS events hours to days after the detection of the GW signal 
when partial disruption of the NS occurs during the merger.

\footnotesize{
\section*{\footnotesize{Acknowledgements}}
We thank F.~Zappa and S.~Bernuzzi for sharing EoS tables.
The authors acknowledge support from INFN, under the Virgo-Prometeo initiative.
O.~S. acknowledges the Italian Ministry for University and Research (MIUR) for funding through project grant 1.05.06.13. 
During drafting of this paper, M.~C. acknowledges kind hospitality by the Kavli Institute for Theoretical Physics at Santa Barbara, under the program "The New Era of Gravitational-Wave Physics and Astrophysics".
}

\footnotesize{
\bibliographystyle{}
\bibliography{references}
}

%
%

\end{document}